\documentstyle[12pt,a4,eepic,epic]{article}
\topmargin - 1.0 true cm
\def\Bsll{B \to X_s l^+ l^-}
\def\Bsg{B \to X_s \gamma}
\renewcommand{\thesection}{\Roman{section}}
\renewcommand{\theequation}{\arabic{section}.\arabic{equation}}
\newcommand{\myequation}[1]{\equation{#1}\setcounter{equation}{1}}

\begin{document}
\baselineskip=7truemm

\begin{flushright}
{\bf hep-ph/0003053}\\ 
HUPD-9920\\
YUMS 00-002\\
KEK-TH-666
\end{flushright}

\begin{center}
\bigskip
{\Large \bf The Effects of Non-Local Interactions \\
            in Rare $B$ Decays, $\Bsll$}
\bigskip\bigskip

{S.~ Fukae$^{~a,}$\footnote{fukae@hiroshima-u.ac.jp},~~~
C.~ S.~ Kim$^{~b,}$\footnote{kim@kimcs.yonsei.ac.kr,
   ~~http://phya.yonsei.ac.kr/\~{}cskim/}~~~and~~~
T.~ Yoshikawa$^{~c,}$\footnote{JSPS Research Fellow,
yosikawa@acorn.kek.jp}}
\end{center}

\bigskip\bigskip

\begin{tabular}{l}
$a$ : {\it Dept. of Physics, Hiroshima Univ.,  Higashi Hiroshima
		739-8526, Japan}\\
$b$ : {\it Depart of Physics and IPAP, Yonsei University, Seoul 120-749,
Korea}\\ $c$ : {\it Theory Group, KEK, Tsukuba, Ibaraki 305-0801, Japan}
\end{tabular}

\bigskip\bigskip
\begin{abstract}

\noindent
The effects of non-local interactions
in rare $B$ decays, $\Bsll$, are investigated. 
We 
show the correlation between the branching ratio and 
the forward-backward asymmetry via two coefficients of 
the non-local interactions.
This will certainly help us find any deviations
from the standard model through the non-local interactions.  

\end{abstract}
\newpage

%
%
\section{Introduction}\label{sec:intro}

Flavor changing neutral current (FCNC) processes are
possibly the most sensitive to the various theoretical extensions of the
standard model (SM) because these
decays occur in the SM only through loops. 
Non-standard model effects can manifest themselves in these rare decays
through the Wilson coefficients, 
which can have values distinctly different from
their standard model counterparts. (See, for example,
\cite{Goto,Wells,handoko,Wyler,Grossman,Rizzo,Jang,Cho,KKL,HHW,LMSS} for the
model dependent analysis.) 
Previously we gave the model-independent analysis on rare $B$ meson decays
$\Bsll$ in Refs. \cite{FKMY,FKY}, 
where we dealt with all the possible local interactions and investigated the
property of these interactions. However, we did not include
the non-local interactions, just for simplicity. These non-local interactions
influence  the process $\Bsg$. 
This process is a kind of ``rare'' decay with
quite large branching ratio, of order of $10^{-4}$ \cite{CLEObsg}, and it
has been studied extensively in Refs. \cite{CMMbsg,CDGG,Misiakbsg,BorGrb}.
Compared to $\Bsg$, the decay $\Bsll$ is much more sensitive to the
actual form of the new interactions since we can measure experimentally
various kinematical distributions as well as the total rate. 
While new physics
can change only the systematically uncertain normalization for $\Bsg$, the
interplay of various operators will change the spectra of the decay
$\Bsll$.  {\it E.g.}, the experimental observation of ${\Bsg}$ 
restricts the absolute value of the Wilson coefficient $C_7^{eff}$ of the
non-local interaction $O_7$,  however, we cannot determine the
sign of the $C_7^{eff}$ from the decay rate of $\Bsg$. But, if we analyze the
interference between the non-local interactions and the other operators 
in the process $\Bsll$, 
we can extract  much more information about $C_7^{eff}$. Therefore, to
search for new physics, it would be most interesting to have a
model-independent study for the non-local interactions in $\Bsll$ decays.

Here we consider the branching ratio and the forward-backward
(FB) asymmetry of inclusive $B\rightarrow X_s e^+ e^-$ or 
$B\rightarrow X_s \mu^+ \mu^- $ decay, 
which are functions of the twelve Wilson
coefficients of four-Fermi interactions. The corresponding matrix elements 
\cite{FKMY,FKY} are given
as
\begin{eqnarray}
{\cal M}(B \to X_s l^+ l^-) &=& 
     \frac{G_F~ \alpha }{\sqrt{2}\pi }~V_{ts}^*V_{tb} \times \nonumber \\
              &[& C_{SL} ~\bar{s} i \sigma_{\mu \nu } \frac{q^\nu}{q^2} 
                        ( m_s L ) b 
                         ~\bar{l} \gamma^\mu l
              + C_{BR} ~\bar{s} i \sigma_{\mu \nu } \frac{q^\nu}{q^2} 
                        ( m_b R ) b 
                        ~\bar{l} \gamma^\mu l  \nonumber \\ 
              &+& C_{LL} ~\bar{s}_L \gamma_\mu b_L 
                           ~\bar{l}_L \gamma^\mu l_L  
              + C_{LR} ~\bar{s}_L \gamma_\mu b_L  
                           ~\bar{l}_R \gamma^\mu l_R  \nonumber \\
              &+& C_{RL} ~\bar{s}_R \gamma_\mu b_R 
                           ~\bar{l}_L \gamma^\mu l_L  
              + C_{RR} ~\bar{s}_R \gamma_\mu b_R  
                           ~\bar{l}_R \gamma^\mu l_R  \nonumber \\ 
              &+& C_{LRLR} ~\bar{s}_L b_R ~\bar{l}_L l_R 
              + C_{RLLR} ~\bar{s}_R b_L ~\bar{l}_L l_R \nonumber \\
              &+& C_{LRRL} ~\bar{s}_L b_R ~\bar{l}_R l_L  
              + C_{RLRL} ~\bar{s}_R b_L ~\bar{l}_R l_L \nonumber \\ 
              &+& C_T      ~\bar{s} \sigma_{\mu \nu } b 
                           ~\bar{l} \sigma^{\mu \nu } l 
              + i C_{TE}   ~\bar{s} \sigma_{\mu \nu } b 
                     ~\bar{l} \sigma_{\alpha \beta } l 
                     ~\epsilon^{\mu \nu \alpha \beta } ].\label{eqn:matrix}
\end{eqnarray}
Here, we represent the Wilson coefficients as $C_{XX}$'s. The $C_{SL}$ and
$C_{BR}$ correspond to the non-local four-Fermi operators, and the other ten
coefficients to the local operators. 
We choose the mass of $b$-quark, $m_b = 4.8$ GeV, 
as the renormalization scale $\mu$. 
The subscripts, $L$ and $R$, express chiral projection operators, 
$L = \frac{1}{2}(1-\gamma_5)$ and $R = \frac{1}{2}(1+\gamma_5)$, and thus
correspond to the chirality of quark and lepton operators. 
Thus, there are  two non-local interactions, $C_{SL}$ and $C_{BR}$ and
ten local ones, {\it i.e.}, four vector-type interactions $C_{LL}$,
$C_{LR}$, $C_{RL}$ and $C_{RR}$, four scalar-type ones 
$C_{LRLR}$, $C_{LRRL}$,
$C_{RLLR}$ and $C_{RLRL}$, and two tensor-type ones $C_T$ and $C_{TE}$.  
We note that two coefficients of the non-local interactions are also constrained
by the experimental data of $\Bsg$, which will be shown in Sec. \ref{sec:decaydistrib}.

The SM predicts that:
\begin{itemize}
\item  
Both of the $C_{SL}$ and $C_{BR}$ are equal to  $-2 C_7^{eff}$, {\it i.e.},
\begin{equation}
m_s^2\left|C_{SL}\right|^2+m_b^2\left|C_{BR}\right|^2
   =4\left(C_7^{eff}\right)^2(m_b^2+m_s^2) \approx 4\left(C_7^{eff}\right)^2 m_b^2.\label{eqn:sm7eff}
\end{equation}
\item  
The $C_{LL}$ and $C_{LR}$ in vector parts are given in
terms of $C_9^{eff}$ and $C_{10}$, that is, 
\begin{equation}
C_{LL} = C_9^{eff} - C_{10}~~~~~ {\rm and}~~~~~ C_{LR} = C_9^{eff} + C_{10}.
\end{equation} 
\item  
The other coefficients are all negligible, and  
${\cal M}(B \to X_s l^+ l^-)_{\rm SM}$ becomes
\begin{eqnarray}
{\cal M}(B \to X_s l^+ l^-)_{\rm SM} &\approx&
	\frac{G_F\alpha}{\sqrt{2}\pi}V_{ts}^*V_{tb} \times \nonumber\\
	&[&-2C_7^{eff}\bar{s}i\sigma_{\mu\nu}\frac{q^{\nu}}{q^2}
	\left(m_sL+m_bR\right)b\bar{l}\gamma^{\mu}l\nonumber\\
&+&\left(C_9^{eff}-C_{10}\right)\bar{s}_L\gamma_{\mu}b_L\bar{l}_L\gamma^{\mu}l_L\nonumber\\
&+&\left(C_9^{eff}+C_{10}\right)\bar{s}_L\gamma_{\mu}b_L\bar{l}_R\gamma^{\mu}l_R] .\label{eqn:matrixsm}
\end{eqnarray}
We incorporate the long distance effects of charmonium states
$J/\psi, ~\psi'$ and higher resonances into the coefficient $C_9^{eff}$,
following Refs. \cite{Long,Kruger}. 
\item  The three coefficients in the SM have been well studied 
\cite{Misiak,BM}, and we follow Ref. \cite{AGHM,KMS} for their choice and set
$$
(C_7^{eff}, C_9^{NDR}, C_{10}) = (-0.311, 4.153, -4.546).
$$
\item 
We finally note that Eq. (\ref{eqn:matrix}) is a model independent
expression \cite{FKMY,FKY} as a whole,   
even though the first two interaction terms (non-local interactions) imply  
the left-handed $b$-quark always comes
with $m_s$ and the right-handed $b$-quark with $m_b$, {\it e.g.} 
the patterns similar to the SUGRA model.  
This is just for a convenient scaling to compare the
Wilson coefficients of the non-local interactions $C_{SL},~C_{BR}$ with
$C_7^{eff}$ of the SM and to get the constraints on them, 
{\it i.e.} the first two terms in Eqs. (\ref{eqn:matrix}) and (\ref{eqn:matrixsm}) compared as;
$$ 
C_{SL}m_s L + C_{BR} m_b R ~~ \Longleftrightarrow  ~~     
-2 C_7^{eff}m_s L - 2 C_7^{eff}m_b R, 
$$
as shown in Figure \ref{fig:brslconstraint} and Eqs. (\ref{eqn:slbrconstraint})
- (\ref{eqn:CSLlimit}). (We assume that interactions due to lepton chirality
flip like $\frac{q_{\nu}}{q^2}\bar{s}\gamma^{\mu}b\bar{l}m_l\sigma_{\mu\nu}l$ is
negligible as $m_l \to 0$.)
\end{itemize}
The prediction of the SM, as shown in Eq.(\ref{eqn:sm7eff}), the interaction
$\bar{s}i\sigma_{\mu\nu}\frac{q^{\nu}}{q^2}(m_sL)b\bar{l}\gamma^{\mu}l$ is
almost negligible in comparison with the other nonlocal interaction. However, in
other models, it is not always so. (In fact, $m_s C_{SL} = m_b C_{BR}$ is the
case in the left-right symmetric model.) One of our aim is that we know how much the
former interaction gives influence to the precess $\Bsll$ in model-independent
way, based on our previous works\cite{FKMY}, where we examined new physics
in the form of {\em local} interaction systematically. We will consider on the
interference between such {\em non-local} interactions and other interactions,
specially the SM interactions, and try to extract the above information. 

The paper is organized as follows. 
In Sec. \ref{sec:decaydistrib}, we 
study the effects due to the non-local interactions on  the branching ratio
and the FB asymmetry, which are derived from the most general
effective Hamiltonian. In Sec.
\ref{sec:behavior}, we  give the correlation between the branching ratio and
the FB asymmetry, which gives very useful information to understand the
interaction from new physics. Conclusions are also in Sec.\ref{sec:behavior}.

%
%
\setcounter{equation}{0}
\section{Branching Ratio and Forward-Backward Asymmetry
of the Process, $\Bsll$}\label{sec:decaydistrib}

We calculate the branching ratio and the forward-backward (FB) asymmetry of 
the $\Bsll$ decay due to the new operators of the models  beyond the SM,
following the method \cite{FKMY,FKY}. We first
concentrate on  the $\Bsg$ decay to get the present constraints on 
the non-local Wilson coefficients of the $\Bsll$ decay.
The effective Hamiltonian for the $\Bsg$ is given as
\begin{equation}
	{\cal M}(\Bsg) = -\frac{4G_F}{\sqrt{2}}V_{ts}^*V_{tb}
		\sum^8_iC_i(\mu){\cal O}_i(\mu)\label{eqn:matrixBsg},
\end{equation}
where $C_i$'s and ${\cal O}_i$'s are the relevant Wilson coefficients and the
corresponding operators. We show only the ${\cal O}_7$ explicitly, that is,
\begin{equation}
	{\cal O}_7 = \frac{e}{16\pi^2}~\bar{s}\sigma_{\mu\nu}
                 (m_b~R + m_s~L)b~F^{\mu\nu},
		\label{eqn:bsgop}
\end{equation}
where $e$ and $F^{\mu\nu}$ are the electromagnetic coupling constant and the
electromagnetic field strength. The resultant branching ratio in the leading
order is given as
\begin{equation}
	{\cal B}(\Bsg) = {\cal B}_0 \frac{32 \pi}{\alpha}~
             \left|C_7^{eff}\right|^2\label{eqn:brBsg},
\end{equation}
where ${\cal B}_0$ is the normalization factor, 
normalized to the semi-leptonic
branching fraction ${\cal B}_{sl}(B\to Xl\nu)$ as
\begin{equation}
{\cal B}_0 = {\cal B}_{sl}~ \frac{3 \alpha^2 }{16 \pi^2 }~
             \frac{ |V_{ts}^*V_{tb}|^2 }{|V_{cb}|^2 }~ 
            \frac{1}{f(\hat{m_c}) \kappa(\hat{m_c})}. 
\label{eqn:semilepfac}
\end{equation}
Here $f(\hat{m_c}={m_c \over m_b})$ and $\kappa(\hat{m_c})$ are phase space
factor and the ${\cal O}(\alpha_s)$ QCD correction factor \cite{Kim} of a
process $b \rightarrow c l \nu$ given by
\begin{eqnarray}
f(\hat{m_c}) &=& 1 - 8 \hat{m_c}^2 + 8 
     \hat{m_c}^6 - \hat{m_c}^8 - 24 \hat{m_c}^4 \ln \hat{m_c} ,  
\label{eqn:phfac}\\
\kappa(\hat{m_c}) &=& 1 - \frac{2 \alpha_s(m_b)}{3 \pi}~ 
           \left[\left(\pi^2-\frac{31}{4}\right)\left(1-\hat{m_c}\right)^2 +
		\frac{3}{2} \right] . \label{eqn:qcdcorr}
\end{eqnarray}
For the numerical analysis, 
we set $\frac{|V_{ts}^*V_{tb}|^2}{|V_{cb}|^2}=1$ and use 
${\cal B}_{sl}=10.4\%$,  the experimental value of semileptonic branching
fraction of $B\to Xl\nu$. By measuring the branching fraction of $\Bsg$, 
we can find present constraints on the theory describing the decay $\Bsll$,
especially on $C_7^{eff}$,  which also appears as the coefficient of the non-local
operators  of the decay $\Bsll$ \cite{Wells}.

\begin{figure}
\vspace*{0.5cm}
\setlength{\unitlength}{0.00053333in}
\begingroup\makeatletter\ifx\SetFigFont\undefined%
\gdef\SetFigFont#1#2#3#4#5{%
  \reset@font\fontsize{#1}{#2pt}%
  \fontfamily{#3}\fontseries{#4}\fontshape{#5}%
  \selectfont}%
\fi\endgroup%
{\renewcommand{\dashlinestretch}{30}
\begin{picture}(4845,4859)(-3000,-200)
\put(8.543,7.543){\whiten\arc{8449.056}{4.7182}{6.2774}}
\put(8.543,7.543){\arc{8449.056}{4.7182}{6.2774}}
\put(48.000,47.000){\whiten\arc{7170.063}{4.7082}{6.2874}}
\put(48.000,47.000){\arc{7170.063}{4.7082}{6.2874}}
\path(33,4832)(33,32)(4833,32)
\put(2033,-400){$m_s\left|C_{SL}\right|$}
\put(-1400,2000){$m_b\left|C_{BR}\right|$}
\put(35,3900){$\times$~SM}
\put(3500,3000){$III$}
\put(3000,2300){$II$}
\put(1000,1000){$I$}
\end{picture}
}
\caption{Constraint on $C_{SL}$ and $C_{BR}$ by rare B decays $B\to
X_s\gamma$. These coefficients can have values only within the region
$II$. The mark $\times$ denotes the standard model point.}
\label{fig:brslconstraint}
\end{figure}
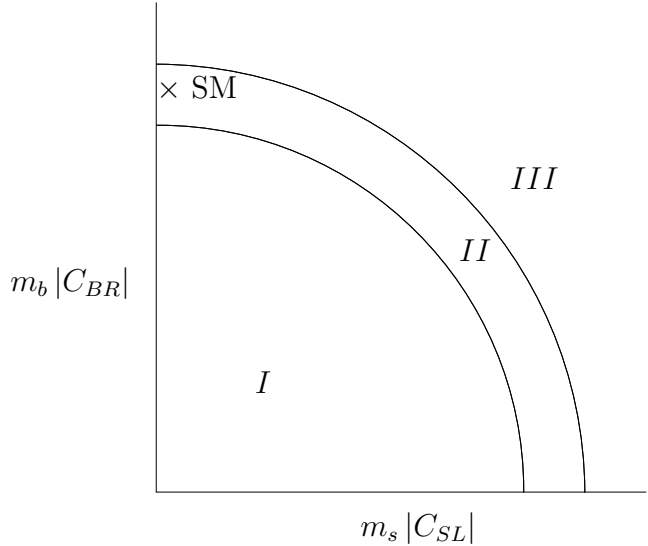

Based on the experimental values of the decay width of $\Bsg$, which is
consistent with the value of $C_7^{eff}$ predicted by the SM \cite{CLEObsg,CDGG}
as appearing in Eq. (\ref{eqn:sm7eff}),
\begin{equation}
	4\left(C_7^{eff}\right)^2(m_b^2+m_s^2)=m_s^2\left|C_{SL}\right|^2
		+m_b^2\left|C_{BR}\right|^2,\label{eqn:slbrconstraint}
\end{equation}
we can easily find that the coefficients of two non-local
operators are placed between two fans whose radii are about 
$2(m_b^2 + m_s^2)^{1/2}|C_7^{eff}|$ in 
$(m_s \left|C_{SL}\right|, m_b \left|C_{BR}\right|)$ plain. 
And the recent result at CLEO for the branching ratio of the $\Bsg$
\cite{CLEObsg},
\[2.0\times 10^{-4} < {\cal B}(\Bsg) < 4.5\times 10^{-4} ~(95\% ~CL),\]
gives the constraint on the absolute value of $C_7^{eff}(m_b)$
\cite{AliHan}, that is,
\[0.28 < |C_7^{eff}(m_b)| < 0.41.\]
As shown in  Figure \ref{fig:brslconstraint}, 
only the values of the coefficients $C_{BR}$ and $C_{SL}$ within
the region $II$  can be permitted.  
Because $s$-quark mass is much less than $b$-quark mass, $m_s \ll m_b$, 
we may regard that the term $C_{SL}$, which is proportional to $m_s$,
hardly contributes to the $\Bsll$ decay in the SM. This means
that the SM point is placed near the $m_b\left|C_{BR}\right|$ axis in the
$(m_s\left|C_{SL}\right|, m_b\left|C_{BR}\right|)$ plain, as shown in Figure
\ref{fig:brslconstraint}. 
Assuming that there is no new phase from the non-local interactions,  
Eq. (\ref{eqn:slbrconstraint}) gives, as $m_s \to 0$,
\begin{equation}
 -2C_7^{eff} \leq C_{BR}\leq 2C_7^{eff},\label{eqn:CBRlimit}
\end{equation}
and
\begin{equation}
 -2C_7^{eff} \leq C_{SL}^N \leq 2C_7^{eff} ,~~~~{\rm where}~~~~
 C_{SL}^N \equiv \frac{m_s}{m_b}C_{SL} . \label{eqn:CSLlimit}
\end{equation}
Here, we denoted the normalized $C_{SL}$ as $C_{SL}^N$.  
Therefore, it is very important to know the branching ratio at 4  points 
$$
(C_{SL}^N,~C_{BR}) = (-2 C_7^{eff}, 0),~~(2 C_7^{eff}, 0),~~(0, -2C_7^{eff}),~~(0, 2C_7^{eff}).
$$ 
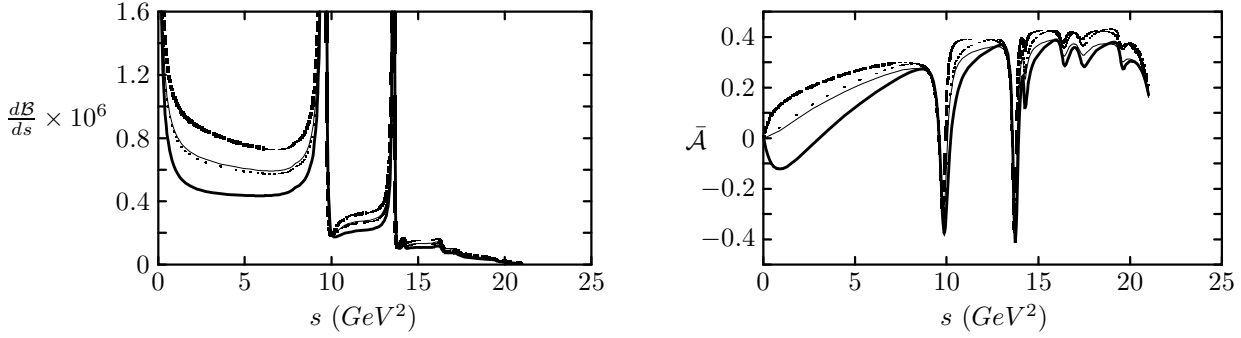
\begin{figure}
\begin{center}
\begin{minipage}[t]{3.0in}
\setlength{\unitlength}{0.14500pt}
\begin{picture}(1500,900)(0,0)
\footnotesize
\thicklines \path(264,90)(284,90)
\thicklines \path(1394,90)(1374,90)
\put(0,470){\makebox(0,0){$\frac{d{\cal B}}{ds}\times 10^6$}}
\put(242,90){\makebox(0,0)[r]{$0$}}
\thicklines \path(264,173)(284,173)
\thicklines \path(1394,173)(1374,173)
\thicklines \path(264,255)(284,255)
\thicklines \path(1394,255)(1374,255)
\put(242,255){\makebox(0,0)[r]{$0.4$}}
\thicklines \path(264,338)(284,338)
\thicklines \path(1394,338)(1374,338)
\thicklines \path(264,420)(284,420)
\thicklines \path(1394,420)(1374,420)
\put(242,420){\makebox(0,0)[r]{$0.8$}}
\thicklines \path(264,503)(284,503)
\thicklines \path(1394,503)(1374,503)
\thicklines \path(264,585)(284,585)
\thicklines \path(1394,585)(1374,585)
\put(242,585){\makebox(0,0)[r]{$1.2$}}
\thicklines \path(264,668)(284,668)
\thicklines \path(1394,668)(1374,668)
\thicklines \path(264,750)(284,750)
\thicklines \path(1394,750)(1374,750)
\put(242,750){\makebox(0,0)[r]{$1.6$}}
\thicklines \path(264,90)(264,110)
\thicklines \path(264,750)(264,730)
\put(800,-50){\makebox(0,0){$s$~($GeV^2$)}}
\put(264,45){\makebox(0,0){$0$}}
\thicklines \path(490,90)(490,110)
\thicklines \path(490,750)(490,730)
\put(490,45){\makebox(0,0){$5$}}
\thicklines \path(716,90)(716,110)
\thicklines \path(716,750)(716,730)
\put(716,45){\makebox(0,0){$10$}}
\thicklines \path(942,90)(942,110)
\thicklines \path(942,750)(942,730)
\put(942,45){\makebox(0,0){$15$}}
\thicklines \path(1168,90)(1168,110)
\thicklines \path(1168,750)(1168,730)
\put(1168,45){\makebox(0,0){$20$}}
\thicklines \path(1394,90)(1394,110)
\thicklines \path(1394,750)(1394,730)
\put(1394,45){\makebox(0,0){$25$}}
\thicklines \path(264,90)(1394,90)(1394,750)(264,750)(264,90)
\put(829,818){\makebox(0,0){$$}}
\Thicklines \path(273,750)(273,735)(275,644)(276,602)(278,570)(280,523)(282,484)(285,453)(287,431)(290,412)(294,386)(300,363)(305,348)(310,336)(315,327)(321,318)(326,312)(336,303)(347,296)(357,291)(366,287)(376,284)(386,281)(405,278)(426,275)(446,273)(464,272)(483,271)(494,270)(504,270)(509,270)(514,270)(516,269)(518,269)(520,269)(523,269)(524,269)(525,269)(526,269)(526,269)(528,269)(529,269)(530,269)(531,269)(532,269)(533,269)(534,269)(535,269)(537,269)(540,269)(543,270)
\Thicklines \path(543,270)(548,270)(554,270)(559,270)(563,271)(572,272)(577,272)(582,273)(587,274)(592,275)(601,278)(607,280)(611,283)(614,284)(617,287)(622,291)(632,296)(636,300)(641,305)(647,312)(652,321)(657,333)(662,349)(665,360)(668,374)(670,390)(672,410)(675,439)(678,476)(679,506)(681,540)(682,578)(684,632)(685,691)(686,750)
\Thicklines \path(704,750)(705,530)(705,446)(706,389)(707,315)(709,265)(710,235)(711,215)(712,199)(713,188)(714,179)(715,175)(716,173)(717,169)(718,167)(718,166)(720,164)(721,163)(722,162)(723,162)(724,162)(725,162)(727,162)(728,162)(730,163)(732,164)(742,168)(751,172)(761,174)(770,176)(780,178)(801,180)(811,181)(820,182)(825,183)(829,185)(835,187)(840,189)(844,192)(847,195)(849,198)(854,205)(856,210)(859,217)(861,228)(862,236)(864,244)(865,255)(866,268)(867,285)(868,296)
\Thicklines \path(868,296)(869,311)(870,344)(871,390)(872,459)(873,564)(874,750)
\Thicklines \path(881,750)(882,373)(883,280)(883,231)(884,183)(886,158)(887,146)(888,139)(889,135)(891,133)(891,133)(892,132)(893,132)(893,132)(894,132)(895,133)(896,133)(897,134)(899,136)(901,139)(903,140)(904,142)(905,143)(906,142)(907,141)(907,139)(909,136)(910,134)(911,133)(912,132)(913,132)(915,132)(916,132)(918,133)(923,134)(928,134)(930,134)(933,135)(935,135)(936,135)(937,135)(939,135)(940,135)(941,135)(942,135)(943,135)(944,135)(946,135)(947,135)(948,135)(948,135)
\Thicklines \path(948,135)(951,135)(953,135)(956,135)(958,135)(960,135)(961,135)(962,135)(963,135)(964,135)(966,135)(967,135)(968,135)(969,135)(970,135)(971,135)(972,135)(974,135)(977,135)(979,135)(981,136)(984,136)(986,137)(991,138)(992,138)(993,139)(994,139)(996,139)(996,139)(997,138)(998,138)(999,137)(1001,135)(1005,127)(1007,126)(1008,124)(1009,123)(1011,122)(1012,122)(1013,121)(1013,121)(1014,121)(1015,121)(1016,121)(1017,121)(1018,121)(1019,121)(1020,121)(1021,121)(1021,121)(1022,121)
\Thicklines \path(1022,121)(1024,121)(1026,121)(1028,121)(1029,121)(1030,121)(1030,121)(1031,121)(1032,121)(1033,121)(1034,121)(1034,121)(1036,121)(1037,121)(1037,120)(1038,120)(1040,120)(1041,120)(1042,119)(1047,116)(1052,114)(1055,112)(1057,112)(1060,111)(1062,110)(1067,109)(1076,108)(1097,106)(1117,104)(1128,103)(1133,103)(1135,103)(1136,103)(1137,103)(1138,102)(1139,102)(1140,102)(1142,102)(1143,102)(1145,101)(1147,99)(1150,99)(1151,98)(1152,98)(1155,97)(1158,97)(1178,95)(1213,92)
\thinlines \path(273,750)(273,734)(275,643)(276,601)(278,570)(280,522)(282,484)(285,453)(287,431)(290,412)(294,386)(300,363)(305,348)(310,335)(315,326)(321,318)(326,311)(336,302)(347,295)(357,291)(366,287)(376,284)(386,281)(405,278)(426,275)(446,273)(464,272)(483,270)(494,270)(504,270)(509,269)(514,269)(516,269)(518,269)(520,269)(523,269)(524,269)(525,269)(526,269)(526,269)(528,269)(529,269)(530,269)(531,269)(532,269)(533,269)(534,269)(535,269)(537,269)(540,269)(543,269)
\thinlines \path(543,269)(548,270)(554,270)(559,270)(563,270)(572,271)(577,272)(582,273)(587,274)(592,275)(601,278)(607,280)(611,282)(614,284)(617,286)(622,291)(632,296)(636,300)(641,305)(647,312)(652,321)(657,332)(662,349)(665,360)(668,374)(670,389)(672,410)(675,439)(678,476)(679,505)(681,540)(682,578)(684,632)(685,691)(686,750)
\thinlines \path(704,750)(705,531)(705,447)(706,389)(707,315)(709,265)(710,235)(711,215)(712,200)(713,188)(714,179)(715,175)(716,173)(717,169)(718,167)(718,166)(720,164)(721,163)(722,162)(723,162)(724,162)(725,162)(727,162)(728,162)(730,163)(732,164)(742,168)(751,172)(761,174)(770,176)(780,178)(801,180)(811,181)(820,182)(825,183)(829,184)(835,186)(840,189)(844,192)(847,195)(849,198)(854,205)(856,210)(859,217)(861,228)(862,235)(864,244)(865,255)(866,268)(867,285)(868,296)
\thinlines \path(868,296)(869,311)(870,344)(871,390)(872,459)(873,564)(874,750)
\thinlines \path(881,750)(882,373)(883,280)(883,231)(884,183)(886,158)(887,146)(888,139)(889,135)(891,133)(891,133)(892,132)(893,132)(893,132)(894,132)(895,133)(896,133)(897,134)(899,136)(901,138)(903,140)(904,142)(905,143)(906,142)(907,141)(907,139)(909,136)(910,134)(911,133)(912,132)(913,132)(915,132)(916,132)(918,133)(923,134)(928,134)(930,134)(933,134)(935,135)(936,135)(937,135)(939,135)(940,135)(941,135)(942,135)(943,135)(944,135)(946,135)(947,135)(948,135)(948,135)
\thinlines \path(948,135)(951,135)(953,135)(956,135)(958,135)(960,135)(961,135)(962,135)(963,135)(964,135)(966,135)(967,135)(968,135)(969,135)(970,135)(971,135)(972,135)(974,135)(977,135)(979,135)(981,136)(984,136)(986,137)(991,138)(992,138)(993,138)(994,139)(996,139)(996,139)(997,138)(998,138)(999,137)(1001,135)(1005,127)(1007,126)(1008,124)(1009,123)(1011,122)(1012,122)(1013,121)(1013,121)(1014,121)(1015,121)(1016,121)(1017,121)(1018,121)(1019,121)(1020,120)(1021,120)(1021,120)(1022,120)
\thinlines \path(1022,120)(1024,121)(1026,121)(1028,121)(1029,121)(1030,121)(1030,121)(1031,121)(1032,121)(1033,121)(1034,121)(1034,121)(1036,121)(1037,120)(1037,120)(1038,120)(1040,120)(1041,119)(1042,119)(1047,116)(1052,114)(1055,112)(1057,112)(1060,111)(1062,110)(1067,109)(1076,108)(1097,106)(1117,104)(1128,103)(1133,103)(1135,103)(1136,103)(1137,103)(1138,102)(1139,102)(1140,102)(1142,102)(1143,102)(1145,101)(1147,99)(1150,99)(1151,98)(1152,98)(1155,97)(1158,97)(1178,95)(1213,92)
\Thicklines \dashline[-20]{1}(278,750)(280,713)(282,674)(285,643)(287,620)(290,600)(294,573)(300,548)(305,532)(315,507)(321,497)(326,488)(337,476)(347,466)(367,452)(385,441)(424,424)(465,410)(504,399)(524,395)(542,392)(546,391)(552,390)(556,390)(561,390)(564,389)(566,389)(568,389)(569,389)(570,389)(570,389)(571,389)(572,389)(573,389)(574,389)(575,389)(577,389)(578,389)(579,389)(582,389)(584,390)(587,390)(591,391)(596,392)(601,393)(606,395)(608,396)(611,398)(613,399)(614,400)(615,401)
\Thicklines \dashline[-20]{1}(615,401)(617,404)(619,407)(624,409)(630,412)(635,416)(639,421)(643,426)(648,434)(653,445)(657,458)(662,478)(667,503)(669,520)(672,544)(674,576)(677,615)(678,641)(679,674)(681,709)(682,750)
\Thicklines \dashline[-20]{1}(703,750)(704,510)(705,363)(706,312)(706,272)(708,224)(709,196)(710,182)(711,172)(713,166)(714,164)(715,163)(716,164)(717,165)(718,166)(719,168)(722,173)(727,182)(732,191)(738,198)(742,203)(748,207)(757,213)(762,215)(767,217)(773,219)(777,220)(787,222)(797,223)(808,224)(813,225)(818,226)(823,227)(828,229)(833,231)(836,232)(839,234)(843,238)(848,243)(850,247)(853,252)(855,258)(857,265)(860,277)(862,292)(864,302)(865,317)(866,336)(868,361)(869,393)(870,437)
\Thicklines \dashline[-20]{1}(870,437)(871,494)(873,591)(874,750)
\Thicklines \dashline[-20]{1}(881,750)(881,423)(882,247)(883,181)(884,155)(886,141)(887,135)(888,133)(889,133)(890,134)(891,136)(896,145)(899,150)(901,156)(903,158)(903,159)(904,160)(905,159)(906,156)(907,152)(908,148)(910,145)(911,144)(912,144)(913,145)(914,145)(916,146)(921,149)(924,150)(926,150)(929,151)(931,151)(934,152)(936,152)(939,152)(941,152)(943,152)(944,153)(945,153)(945,153)(946,153)(947,153)(948,153)(949,153)(950,153)(951,153)(952,153)(953,153)(954,153)(957,153)(959,153)
\Thicklines \dashline[-20]{1}(959,153)(960,153)(961,153)(962,153)(964,153)(965,153)(965,153)(966,153)(967,153)(968,153)(970,153)(971,153)(973,153)(974,153)(975,153)(977,153)(980,153)(982,154)(985,154)(989,156)(990,156)(991,156)(993,157)(994,157)(995,157)(996,156)(997,155)(998,155)(999,154)(1001,149)(1003,143)(1005,140)(1006,137)(1008,135)(1009,134)(1010,133)(1012,132)(1012,132)(1013,132)(1014,131)(1015,131)(1015,131)(1017,131)(1018,131)(1019,131)(1019,131)(1021,131)(1022,131)(1024,131)(1026,132)(1027,132)
\Thicklines
\dashline[-20]{1}(1027,132)(1028,132)(1028,132)(1029,132)(1030,132)(1031,132)(1032,132)(1032,132)(1034,132)(1035,131)(1035,131)(1036,131)(1038,131)(1039,130)(1040,130)(1042,128)(1045,127)(1050,122)(1052,121)(1055,119)(1058,118)(1060,117)(1063,117)(1065,116)(1075,115)(1093,112)(1113,109)(1122,108)(1131,107)(1134,107)(1135,107)(1137,107)(1138,107)(1139,106)(1140,106)(1142,106)(1143,105)(1144,104)(1146,103)(1149,101)(1151,100)(1152,100)(1154,100)(1156,99)(1161,99)(1172,97)(1211,92)(1213,92)
\thinlines \path(275,750)(275,743)(276,701)(278,669)(280,622)(282,583)(285,552)(287,530)(290,510)(294,483)(300,460)(305,444)(310,431)(315,421)(326,404)(336,394)(347,385)(368,373)(388,365)(426,353)(467,345)(487,341)(506,339)(524,336)(535,335)(539,335)(544,335)(549,335)(551,334)(553,334)(554,334)(555,334)(556,334)(557,334)(558,334)(559,334)(561,334)(562,334)(563,334)(564,334)(566,334)(568,335)(571,335)(573,335)(578,335)(581,336)(584,336)(588,337)(593,338)(598,339)(603,341)
\thinlines \path(603,341)(608,343)(613,346)(623,355)(628,358)(633,361)(638,366)(643,372)(648,380)(653,390)(658,404)(661,415)(664,428)(669,456)(671,477)(674,502)(676,529)(678,568)(680,624)(683,696)(684,750)
\thinlines \path(703,750)(704,525)(705,432)(706,370)(707,294)(708,245)(709,217)(710,200)(711,187)(713,178)(714,172)(715,170)(715,169)(717,167)(717,166)(718,166)(719,166)(720,167)(722,168)(723,169)(732,178)(741,186)(746,190)(751,193)(760,197)(765,198)(770,200)(780,202)(790,203)(800,204)(811,206)(816,206)(821,207)(826,209)(830,210)(835,212)(840,215)(843,218)(846,220)(848,224)(851,228)(853,233)(856,240)(859,248)(861,259)(862,266)(863,276)(865,286)(866,300)(867,321)(868,334)
\thinlines \path(868,334)(869,350)(869,367)(870,390)(871,445)(872,539)(874,692)(874,750)
\thinlines \path(881,750)(881,542)(882,283)(883,199)(884,166)(886,148)(887,139)(887,137)(888,135)(889,134)(891,134)(892,135)(893,136)(895,138)(899,145)(902,150)(903,152)(904,153)(905,152)(906,150)(908,145)(909,142)(910,140)(911,140)(912,140)(914,140)(915,140)(916,141)(919,142)(921,143)(924,143)(928,144)(930,144)(932,145)(935,145)(937,145)(940,145)(941,145)(942,145)(944,145)(945,145)(947,145)(948,145)(949,145)(949,145)(951,145)(951,145)(952,145)(953,145)(954,145)(957,145)
\thinlines \path(957,145)(958,145)(959,145)(960,145)(960,145)(962,145)(963,145)(964,145)(965,145)(966,145)(967,146)(968,146)(969,146)(970,146)(973,146)(975,146)(977,146)(980,146)(982,147)(985,147)(989,149)(991,149)(992,149)(993,150)(994,150)(995,150)(996,149)(998,148)(998,148)(999,147)(1001,143)(1004,138)(1006,133)(1008,131)(1009,130)(1010,129)(1011,129)(1011,128)(1012,128)(1013,128)(1013,128)(1015,127)(1016,127)(1017,127)(1018,127)(1019,127)(1020,127)(1021,127)(1022,127)(1025,128)(1027,128)
\thinlines \path(1027,128)(1028,128)(1030,128)(1031,128)(1032,128)(1033,128)(1034,128)(1035,127)(1036,127)(1037,127)(1038,127)(1039,127)(1040,126)(1042,126)(1045,124)(1049,120)(1052,118)(1055,117)(1057,116)(1060,115)(1062,114)(1064,114)(1074,113)(1094,110)(1114,108)(1124,107)(1133,106)(1135,106)(1137,106)(1138,106)(1139,105)(1140,105)(1141,105)(1143,104)(1147,102)(1149,101)(1150,100)(1151,100)(1153,99)(1154,99)(1156,99)(1162,98)(1171,97)(1209,93)(1213,92)
\thinlines \dashline[-10]{5}(275,750)(275,734)(276,693)(278,661)(280,614)(282,575)(285,544)(287,521)(290,502)(294,475)(300,452)(305,436)(310,422)(315,412)(326,396)(336,385)(347,377)(368,364)(388,356)(426,345)(467,336)(487,333)(506,330)(524,328)(535,327)(544,326)(549,326)(551,326)(554,326)(556,325)(557,325)(558,325)(559,325)(561,325)(562,325)(563,325)(564,325)(565,325)(566,325)(567,325)(568,326)(570,326)(573,326)(576,326)(578,326)(584,327)(589,328)(594,329)(599,330)(605,332)(607,333)
\thinlines \dashline[-10]{5}(607,333)(610,335)(615,338)(625,346)(630,348)(635,352)(640,357)(645,363)(650,371)(655,382)(660,398)(663,410)(666,423)(670,454)(673,478)(675,506)(677,538)(679,585)(681,615)(682,653)(684,745)(684,750)
\thinlines \dashline[-10]{5}(703,750)(703,749)(705,497)(706,367)(707,320)(707,286)(708,244)(710,215)(711,198)(712,187)(713,179)(714,173)(715,169)(716,168)(717,167)(718,166)(719,166)(719,166)(721,166)(722,167)(723,168)(724,169)(733,176)(742,184)(748,187)(752,189)(761,192)(766,194)(772,195)(782,197)(801,199)(811,200)(817,201)(822,202)(827,203)(832,205)(835,206)(837,207)(842,211)(845,213)(848,216)(850,219)(852,223)(855,228)(857,235)(859,243)(861,253)(863,261)(864,271)(865,282)(866,297)
\thinlines \dashline[-10]{5}(866,297)(867,309)(868,321)(868,335)(869,353)(870,396)(871,428)(872,466)(873,584)(874,750)
\thinlines \dashline[-10]{5}(881,750)(882,395)(883,247)(884,184)(885,158)(886,146)(887,139)(888,136)(889,134)(890,134)(892,134)(893,135)(894,136)(895,137)(900,143)(902,147)(903,148)(904,149)(905,149)(906,147)(908,144)(909,140)(909,139)(910,138)(911,138)(912,137)(913,137)(915,138)(919,139)(922,140)(925,140)(928,141)(930,141)(933,141)(935,142)(936,142)(938,142)(939,142)(940,142)(941,142)(942,142)(943,142)(943,142)(944,142)(945,142)(947,142)(948,142)(949,142)(950,142)(953,142)(956,142)
\thinlines \dashline[-10]{5}(956,142)(959,142)(960,142)(961,142)(962,142)(964,142)(964,142)(965,142)(966,142)(967,142)(968,142)(970,142)(971,142)(972,142)(974,142)(975,142)(976,142)(979,142)(981,143)(984,143)(986,144)(991,145)(992,145)(993,145)(994,146)(995,145)(997,145)(997,145)(998,144)(999,143)(1001,141)(1005,133)(1006,131)(1007,129)(1008,128)(1009,127)(1011,126)(1012,126)(1013,125)(1014,125)(1014,125)(1015,125)(1017,125)(1018,124)(1019,124)(1020,124)(1021,124)(1022,124)(1023,124)(1024,125)(1026,125)
\thinlines
\dashline[-10]{5}(1026,125)(1027,125)(1028,125)(1029,125)(1030,125)(1031,125)(1032,125)(1033,125)(1034,125)(1035,124)(1036,124)(1037,124)(1039,124)(1040,123)(1041,123)(1043,122)(1048,119)(1050,117)(1053,116)(1055,115)(1057,114)(1059,113)(1062,112)(1067,112)(1077,110)(1096,108)(1114,106)(1124,105)(1129,104)(1134,104)(1136,104)(1137,104)(1138,104)(1139,103)(1139,103)(1140,103)(1141,103)(1143,102)(1145,101)(1148,100)(1149,99)(1151,99)(1153,98)(1154,98)(1155,98)(1160,97)(1166,97)(1175,96)(1213,92)
\end{picture}
\end{minipage}
\hfill 
\begin{minipage}[t]{3.0in}
\setlength{\unitlength}{0.14500pt}
\begin{picture}(1500,900)(0,0)
\footnotesize
\thicklines \path(198,90)(218,90)
\thicklines \path(1394,90)(1374,90)
\thicklines \path(198,156)(218,156)
\thicklines \path(1394,156)(1374,156)
\put(176,156){\makebox(0,0)[r]{$-0.4$}}
\thicklines \path(198,222)(218,222)
\thicklines \path(1394,222)(1374,222)
\thicklines \path(198,288)(218,288)
\thicklines \path(1394,288)(1374,288)
\put(176,288){\makebox(0,0)[r]{$-0.2$}}
\thicklines \path(198,354)(218,354)
\thicklines \path(1394,354)(1374,354)
\thicklines \path(198,420)(218,420)
\thicklines \path(1394,420)(1374,420)
\put(176,420){\makebox(0,0)[r]{$0$}}
\put(50,420){\makebox(0,0)[r]{$\bar{{\cal A}}$}}
\thicklines \path(198,486)(218,486)
\thicklines \path(1394,486)(1374,486)
\thicklines \path(198,552)(218,552)
\thicklines \path(1394,552)(1374,552)
\put(176,552){\makebox(0,0)[r]{$0.2$}}
\thicklines \path(198,618)(218,618)
\thicklines \path(1394,618)(1374,618)
\thicklines \path(198,684)(218,684)
\thicklines \path(1394,684)(1374,684)
\put(176,684){\makebox(0,0)[r]{$0.4$}}
\thicklines \path(198,750)(218,750)
\thicklines \path(1394,750)(1374,750)
\thicklines \path(198,90)(198,110)
\thicklines \path(198,750)(198,730)
\put(198,45){\makebox(0,0){$0$}}
\thicklines \path(437,90)(437,110)
\thicklines \path(437,750)(437,730)
\put(437,45){\makebox(0,0){$5$}}
\thicklines \path(676,90)(676,110)
\thicklines \path(676,750)(676,730)
\put(676,45){\makebox(0,0){$10$}}
\put(800,-50){\makebox(0,0){$s$ ($GeV^2$)}}
\thicklines \path(916,90)(916,110)
\thicklines \path(916,750)(916,730)
\put(916,45){\makebox(0,0){$15$}}
\thicklines \path(1155,90)(1155,110)
\thicklines \path(1155,750)(1155,730)
\put(1155,45){\makebox(0,0){$20$}}
\thicklines \path(1394,90)(1394,110)
\thicklines \path(1394,750)(1394,730)
\put(1394,45){\makebox(0,0){$25$}}
\thicklines \path(198,90)(1394,90)(1394,750)(198,750)(198,90)
\put(796,818){\makebox(0,0){$$}}
\thicklines \dashline[-20]{1}(200,423)(200,423)(203,435)(205,445)(210,458)(215,471)(220,479)(231,494)(241,504)(262,521)(285,535)(328,556)(370,572)(410,586)(452,598)(493,608)(514,612)(525,613)(536,615)(542,616)(548,616)(553,616)(555,617)(557,617)(558,617)(559,617)(560,617)(561,617)(563,617)(564,617)(565,617)(566,617)(568,617)(569,617)(570,616)(571,616)(573,616)(574,616)(575,616)(576,616)(577,616)(578,616)(579,616)(580,616)(581,616)(582,616)(583,616)(584,615)(585,615)(586,615)(589,615)
\thicklines \dashline[-20]{1}(589,615)(592,615)(595,614)(598,614)(600,613)(606,612)(609,611)(611,609)(616,606)(622,602)(627,596)(632,589)(634,584)(637,577)(640,569)(642,559)(645,548)(647,533)(650,512)(652,487)(655,455)(657,412)(662,308)(663,274)(664,257)(664,242)(666,222)(667,218)(668,229)(669,254)(670,293)(672,342)(675,451)(676,494)(677,528)(680,574)(681,592)(682,607)(685,626)(686,634)(687,640)(689,649)(692,656)(694,661)(697,665)(699,668)(702,670)(705,672)(707,673)(710,674)(713,675)
\thicklines \dashline[-20]{1}(713,675)(716,675)(719,676)(724,676)(727,677)(730,677)(735,677)(740,678)(745,678)(751,678)(754,678)(756,678)(759,678)(760,678)(761,678)(763,678)(764,678)(765,678)(766,678)(767,678)(768,678)(769,678)(770,678)(772,678)(774,678)(777,678)(779,678)(782,678)(785,677)(788,677)(794,676)(797,676)(799,675)(804,674)(807,673)(810,672)(815,669)(817,667)(820,665)(822,662)(825,659)(827,655)(829,650)(832,642)(833,637)(835,631)(838,616)(839,606)(840,594)(842,579)(843,562)
\thicklines \dashline[-20]{1}(843,562)(845,517)(846,482)(847,440)(850,311)(851,239)(853,181)(854,173)(855,218)(857,309)(858,413)(860,492)(861,547)(862,570)(862,591)(864,619)(864,631)(865,642)(867,657)(868,667)(869,673)(871,678)(872,680)(872,681)(873,680)(875,677)(876,670)(877,659)(879,648)(879,645)(880,645)(881,650)(883,657)(884,664)(885,670)(887,678)(889,681)(890,684)(892,688)(895,691)(897,693)(900,695)(903,696)(906,697)(909,698)(912,699)(914,699)(917,700)(919,700)(922,700)(924,701)
\thicklines \dashline[-20]{1}(924,701)(927,701)(928,701)(930,701)(931,701)(932,701)(934,701)(935,701)(936,701)(937,701)(939,701)(940,701)(942,701)(944,700)(947,700)(950,700)(952,699)(954,698)(957,697)(960,696)(963,694)(965,691)(967,688)(970,685)(975,674)(977,671)(977,670)(978,669)(979,667)(981,667)(982,668)(983,669)(983,670)(986,678)(989,684)(992,689)(993,691)(994,692)(996,694)(997,695)(998,696)(1000,696)(1001,697)(1002,697)(1003,697)(1004,697)(1005,697)(1006,696)(1007,696)(1008,696)(1009,695)
\thicklines \dashline[-20]{1}(1009,695)(1010,695)(1013,693)(1015,691)(1018,688)(1023,682)(1024,680)(1025,679)(1026,679)(1027,678)(1028,678)(1029,678)(1030,678)(1031,678)(1031,678)(1033,679)(1034,680)(1040,686)(1045,691)(1048,693)(1050,695)(1055,697)(1057,698)(1060,698)(1063,699)(1064,699)(1066,700)(1067,700)(1069,700)(1070,700)(1072,700)(1073,700)(1074,700)(1075,700)(1077,700)(1078,700)(1079,700)(1079,700)(1081,700)(1082,700)(1085,700)(1087,699)(1092,699)(1098,698)(1103,696)(1106,695)(1108,694)(1111,693)(1113,691)(1116,689)(1118,687)
\thicklines \dashline[-20]{1}(1118,687)(1121,684)(1123,680)(1128,668)(1129,664)(1131,661)(1131,660)(1132,659)(1133,658)(1134,658)(1135,658)(1136,659)(1137,660)(1138,661)(1139,662)(1140,664)(1142,665)(1143,665)(1144,666)(1145,667)(1146,667)(1148,667)(1149,667)(1149,667)(1150,667)(1152,666)(1153,666)(1154,665)(1159,662)(1163,657)(1173,645)(1179,635)(1184,624)(1193,596)(1198,576)(1203,546)
\Thicklines \path(200,417)(200,417)(205,392)(210,376)(215,363)(218,357)(220,353)(223,349)(225,346)(228,344)(231,342)(232,342)(233,341)(235,341)(236,340)(236,340)(237,340)(239,340)(240,340)(241,340)(242,340)(243,340)(243,340)(244,340)(245,340)(247,341)(250,341)(253,343)(259,345)(264,349)(285,365)(326,403)(369,443)(410,479)(450,510)(492,540)(533,566)(553,578)(576,590)(586,594)(592,596)(598,598)(603,599)(606,600)(607,600)(608,600)(609,601)(610,601)(612,601)(613,601)(614,601)
\Thicklines \path(614,601)(615,601)(617,601)(618,601)(619,600)(621,600)(622,599)(623,599)(625,598)(626,597)(629,595)(632,592)(634,588)(637,583)(639,576)(642,568)(644,556)(647,540)(650,519)(652,495)(654,467)(659,377)(664,258)(665,226)(667,201)(668,185)(669,174)(670,172)(672,177)(673,190)(674,206)(679,289)(684,380)(689,444)(692,470)(695,493)(701,525)(706,546)(711,561)(716,576)(721,586)(732,603)(744,616)(763,634)(785,648)(796,655)(806,660)(811,662)(813,663)(815,663)(816,664)
\Thicklines \path(816,664)(817,664)(817,664)(819,664)(820,664)(821,664)(822,664)(823,664)(824,663)(825,662)(827,661)(828,660)(829,659)(830,657)(833,651)(834,647)(836,641)(837,634)(838,626)(839,615)(841,599)(842,578)(844,551)(845,519)(846,483)(849,372)(851,261)(852,205)(854,166)(855,147)(856,156)(857,184)(859,227)(863,403)(866,474)(868,522)(870,542)(871,561)(872,574)(874,581)(875,582)(876,574)(877,553)(878,526)(880,505)(880,500)(881,500)(883,507)(884,518)(887,545)(889,565)
\Thicklines \path(889,565)(892,580)(895,593)(900,610)(906,624)(912,633)(922,647)(931,657)(942,666)(948,670)(953,674)(956,675)(958,676)(959,676)(960,677)(961,677)(962,677)(963,677)(964,676)(965,676)(966,675)(967,674)(968,672)(970,670)(971,666)(973,661)(974,654)(979,629)(980,622)(981,616)(982,612)(984,610)(985,609)(986,610)(986,610)(988,613)(989,616)(994,632)(999,644)(1001,649)(1004,653)(1005,655)(1007,656)(1008,657)(1009,657)(1010,657)(1011,657)(1011,657)(1012,657)(1013,656)(1014,655)
\Thicklines \path(1014,655)(1015,655)(1016,653)(1017,651)(1019,646)(1024,630)(1027,621)(1028,616)(1029,613)(1031,611)(1032,609)(1033,608)(1034,607)(1036,607)(1037,608)(1038,609)(1039,610)(1045,619)(1055,634)(1061,642)(1066,648)(1077,656)(1087,662)(1098,667)(1102,668)(1107,670)(1108,670)(1110,670)(1111,671)(1112,671)(1113,671)(1115,671)(1116,670)(1117,670)(1118,669)(1120,668)(1121,666)(1122,665)(1122,664)(1124,660)(1125,655)(1126,649)(1128,642)(1130,627)(1131,618)(1133,611)(1134,606)(1135,603)(1136,601)(1137,601)(1139,602)
\Thicklines \path(1139,602)(1140,604)(1142,608)(1145,613)(1148,616)(1151,619)(1152,620)(1154,621)(1155,621)(1156,621)(1158,621)(1159,622)(1159,622)(1161,621)(1162,621)(1163,621)(1164,620)(1167,619)(1170,617)(1174,613)(1179,606)(1185,597)(1190,585)(1195,568)(1198,555)(1201,541)(1203,533)
\thinlines \path(200,420)(200,420)(241,440)(285,470)(327,497)(367,519)(410,541)(451,560)(494,577)(516,585)(536,591)(556,597)(561,598)(566,599)(572,599)(577,600)(582,601)(587,602)(591,602)(592,602)(593,602)(595,602)(596,602)(596,602)(598,602)(598,602)(599,602)(600,602)(601,602)(602,602)(603,602)(604,602)(607,602)(608,601)(609,601)(612,601)(615,600)(617,599)(620,597)(623,595)(625,593)(630,588)(633,584)(635,579)(638,573)(641,564)(643,555)(646,543)(648,528)(650,510)(653,484)
\thinlines \path(653,484)(655,446)(658,399)(660,344)(663,277)(664,242)(666,216)(666,205)(667,197)(668,193)(669,194)(669,198)(670,207)(671,231)(677,363)(679,427)(682,474)(684,506)(687,532)(690,554)(692,569)(697,589)(699,597)(702,604)(707,615)(712,622)(717,628)(722,632)(734,640)(744,645)(764,652)(775,655)(786,657)(791,658)(794,658)(797,659)(799,659)(801,659)(802,659)(803,659)(805,659)(806,659)(807,659)(808,659)(810,659)(811,659)(812,659)(814,658)(815,658)(817,658)(819,657)
\thinlines \path(819,657)(821,655)(824,653)(827,650)(829,646)(832,641)(834,634)(836,624)(837,618)(839,608)(840,596)(841,582)(844,546)(845,518)(846,485)(849,375)(850,299)(852,216)(853,165)(854,153)(855,155)(856,189)(857,253)(860,387)(861,446)(863,489)(865,545)(866,567)(868,586)(869,602)(870,615)(871,620)(872,625)(873,628)(874,629)(874,629)(875,627)(876,618)(877,609)(878,599)(878,589)(879,580)(881,574)(882,579)(883,588)(884,596)(887,613)(890,625)(892,633)(895,639)(900,649)
\thinlines \path(900,649)(903,653)(906,656)(911,660)(921,667)(926,669)(932,671)(937,673)(942,674)(944,675)(947,676)(949,676)(950,676)(951,676)(952,676)(953,676)(954,676)(955,676)(957,676)(958,676)(959,676)(960,676)(961,675)(963,675)(964,674)(965,673)(967,672)(968,670)(969,668)(972,662)(977,647)(978,642)(980,638)(981,636)(982,635)(983,634)(985,635)(986,637)(987,639)(992,650)(994,655)(996,659)(999,662)(1000,663)(1002,664)(1003,665)(1004,666)(1005,666)(1006,666)(1008,667)(1009,666)
\thinlines \path(1009,666)(1010,666)(1011,666)(1013,665)(1014,664)(1015,663)(1017,661)(1022,652)(1024,647)(1027,642)(1029,639)(1029,639)(1030,638)(1031,637)(1031,637)(1032,637)(1033,637)(1034,637)(1035,637)(1036,638)(1037,639)(1038,640)(1043,645)(1049,651)(1054,656)(1058,659)(1063,661)(1069,663)(1073,665)(1078,666)(1081,666)(1084,666)(1086,667)(1087,667)(1089,667)(1090,667)(1091,667)(1092,667)(1093,667)(1095,667)(1096,667)(1096,667)(1098,667)(1099,667)(1100,667)(1101,667)(1104,666)(1105,666)(1107,666)(1109,665)(1112,665)
\thinlines \path(1112,665)(1114,664)(1115,663)(1117,662)(1118,661)(1121,658)(1122,656)(1123,653)(1126,645)(1129,635)(1131,629)(1132,625)(1133,622)(1134,620)(1135,619)(1137,619)(1138,620)(1139,621)(1142,623)(1144,625)(1145,626)(1146,626)(1147,627)(1149,627)(1150,627)(1151,627)(1153,627)(1154,627)(1155,627)(1156,626)(1159,625)(1162,623)(1166,619)(1171,614)(1177,606)(1182,597)(1187,586)(1192,572)(1196,557)(1203,526)
\thinlines \dashline[-10]{5}(200,420)(200,420)(241,447)(285,480)(327,507)(367,530)(410,553)(451,572)(494,589)(516,597)(536,603)(556,609)(561,610)(566,611)(567,611)(569,611)(572,611)(577,612)(582,612)(587,613)(590,613)(591,613)(592,613)(593,613)(594,613)(595,613)(596,613)(598,613)(599,613)(600,613)(601,613)(603,613)(605,613)(608,612)(610,612)(613,611)(617,609)(620,608)(622,606)(627,601)(630,597)(633,592)(638,580)(641,571)(644,558)(646,545)(649,525)(652,497)(655,457)(658,405)(660,346)
\thinlines \dashline[-10]{5}(660,346)(663,283)(664,254)(665,232)(667,215)(667,210)(668,209)(669,215)(671,232)(672,258)(673,285)(676,353)(680,458)(683,503)(686,534)(689,560)(692,581)(694,594)(697,604)(699,614)(702,621)(707,631)(713,639)(718,645)(723,650)(734,657)(743,662)(763,669)(775,673)(785,675)(790,676)(793,676)(795,677)(798,677)(799,677)(800,677)(801,677)(803,677)(804,677)(805,677)(806,677)(807,677)(808,677)(809,677)(810,677)(812,677)(813,676)(815,676)(817,675)(818,674)(821,673)
\thinlines \dashline[-10]{5}(821,673)(823,670)(825,668)(827,666)(829,662)(831,656)(834,648)(836,637)(837,630)(839,620)(840,608)(841,591)(843,572)(844,550)(846,491)(849,383)(850,310)(852,234)(853,186)(854,174)(855,174)(856,202)(857,260)(859,379)(862,481)(863,524)(865,555)(867,599)(869,617)(870,631)(871,636)(872,641)(873,647)(874,648)(875,647)(876,638)(877,628)(878,617)(878,605)(879,596)(880,588)(882,592)(883,601)(884,610)(887,629)(890,643)(892,652)(895,660)(900,669)(903,673)(905,677)
\thinlines \dashline[-10]{5}(905,677)(911,682)(922,690)(928,693)(933,696)(939,698)(943,699)(946,700)(949,701)(950,701)(951,701)(952,701)(953,701)(954,701)(954,701)(956,701)(956,701)(957,701)(958,701)(959,701)(959,701)(960,701)(961,700)(963,700)(964,699)(965,697)(967,696)(968,694)(971,689)(972,684)(974,679)(977,668)(978,663)(979,659)(980,657)(981,656)(982,654)(983,655)(985,656)(986,658)(987,661)(990,669)(993,676)(996,682)(998,686)(1001,690)(1002,691)(1004,692)(1005,692)(1006,693)(1007,693)
\thinlines \dashline[-10]{5}(1007,693)(1009,693)(1010,693)(1011,692)(1012,692)(1013,691)(1016,688)(1019,683)(1024,672)(1027,666)(1028,664)(1029,662)(1030,661)(1031,661)(1032,660)(1033,660)(1033,661)(1034,661)(1035,661)(1037,663)(1038,664)(1041,668)(1051,681)(1056,687)(1062,691)(1067,694)(1072,697)(1081,700)(1087,702)(1092,703)(1095,704)(1097,704)(1099,704)(1100,704)(1102,704)(1103,705)(1103,705)(1105,705)(1106,705)(1107,704)(1109,704)(1110,704)(1111,704)(1112,704)(1114,703)(1115,703)(1116,702)(1118,701)(1119,700)(1120,698)(1122,695)
\thinlines \dashline[-10]{5}(1122,695)(1123,692)(1125,688)(1130,665)(1131,659)(1132,656)(1133,654)(1134,652)(1135,651)(1136,651)(1137,652)(1138,652)(1139,653)(1141,657)(1144,661)(1146,664)(1148,666)(1150,666)(1151,667)(1152,667)(1153,668)(1154,668)(1156,668)(1157,668)(1158,667)(1159,667)(1161,666)(1163,665)(1166,663)(1170,659)(1175,653)(1179,646)(1183,638)(1188,626)(1193,609)(1198,587)(1203,558)
\end{picture}
\end{minipage}
\end{center}
\caption{Branching ratio and FB asymmetry 
for $(C_{SL}^N, C_{BR}) = (0, -2C_7^{eff})$ (thick
solid line), $(0, 2C_7^{eff})$ (thick dashed line), 
$(-2C_7^{eff}, 0)$ (thin solid line) and
$(2C_7^{eff}, 0)$ (dotted line). 
The thick solid line also represents for the
SM.}\label{fig:brratio} 
\end{figure}

\begin{figure}
\begin{center}
\begin{minipage}[t]{3.0in}
\setlength{\unitlength}{0.120900pt}
\begin{picture}(1500,900)(0,0)
\footnotesize
\thicklines \path(264,90)(284,90)
\thicklines \path(1350,90)(1330,90)
\put(242,90){\makebox(0,0)[r]{$3.0$}}
\thicklines \path(264,222)(284,222)
\thicklines \path(1350,222)(1330,222)
\thicklines \path(264,355)(284,355)
\thicklines \path(1350,355)(1330,355)
\put(242,355){\makebox(0,0)[r]{$4.0$}}
\put(-50,488){\makebox(0,0)[l]{${\cal B}\times 10^6$}}
\thicklines \path(264,488)(284,488)
\thicklines \path(1350,488)(1330,488)
\thicklines \path(264,620)(284,620)
\thicklines \path(1350,620)(1330,620)
\put(242,620){\makebox(0,0)[r]{$5.0$}}
\thicklines \path(264,753)(284,753)
\thicklines \path(1350,753)(1330,753)
\thicklines \path(264,885)(284,885)
\thicklines \path(1350,885)(1330,885)
\put(242,885){\makebox(0,0)[r]{$6.0$}}
\thicklines \path(264,90)(264,110)
\thicklines \path(264,885)(264,865)
\put(264,45){\makebox(0,0){$0.0$}}
\thicklines \path(481,90)(481,110)
\thicklines \path(481,885)(481,865)
\put(481,45){\makebox(0,0){$0.2$}}
\thicklines \path(698,90)(698,110)
\thicklines \path(698,885)(698,865)
\put(698,45){\makebox(0,0){$0.4$}}
\put(807,-50){\makebox(0,0){$\theta/(2\pi)$}}
\thicklines \path(916,90)(916,110)
\thicklines \path(916,885)(916,865)
\put(916,45){\makebox(0,0){$0.6$}}
\thicklines \path(1133,90)(1133,110)
\thicklines \path(1133,885)(1133,865)
\put(1133,45){\makebox(0,0){$0.8$}}
\thicklines \path(1350,90)(1350,110)
\thicklines \path(1350,885)(1350,865)
\put(1350,45){\makebox(0,0){$1.0$}}
\thicklines \path(264,90)(1350,90)(1350,885)(264,885)(264,90)
\thinlines \path(264,125)(264,125)(267,125)(269,126)(274,127)(280,128)(285,130)(297,134)(308,140)(331,156)(356,179)(403,237)(448,309)(491,387)(537,475)(581,559)(628,639)(673,704)(696,730)(717,749)(739,765)(751,772)(757,774)(763,777)(768,778)(774,780)(777,780)(780,781)(783,781)(786,781)(788,782)(790,782)(791,782)(793,782)(794,782)(795,782)(797,782)(798,782)(799,782)(802,781)(804,781)(807,781)(813,780)(819,778)(831,774)(843,769)(854,762)(877,745)(898,725)(942,670)(989,596)
\thinlines \path(989,596)(1034,516)(1078,434)(1124,349)(1169,274)(1216,207)(1240,179)(1262,159)(1284,143)(1294,137)(1306,132)(1312,130)(1317,128)(1322,127)(1327,126)(1330,126)(1333,125)(1334,125)(1336,125)(1337,125)(1338,125)(1340,125)(1340,125)(1341,125)(1342,125)(1343,125)(1344,125)(1346,125)(1347,125)(1348,125)(1350,125)
\end{picture}
\end{minipage}
\hfill 
\begin{minipage}[t]{3.0in}
\setlength{\unitlength}{0.120900pt}
\begin{picture}(1500,900)(0,0)
\footnotesize
\thicklines \path(198,90)(218,90)
\thicklines \path(1350,90)(1330,90)
\thicklines \path(198,170)(218,170)
\thicklines \path(1350,170)(1330,170)
\put(176,170){\makebox(0,0)[r]{$0.08$}}
\thicklines \path(198,249)(218,249)
\thicklines \path(1350,249)(1330,249)
\thicklines \path(198,329)(218,329)
\thicklines \path(1350,329)(1330,329)
\put(176,329){\makebox(0,0)[r]{$0.12$}}
\thicklines \path(198,408)(218,408)
\thicklines \path(1350,408)(1330,408)
\thicklines \path(198,488)(218,488)
\thicklines \path(1350,488)(1330,488)
\put(176,488){\makebox(0,0)[r]{$0.16$}}
\put(-50,558){\makebox(0,0)[l]{$\bar{{\cal A}}$}}
\thicklines \path(198,567)(218,567)
\thicklines \path(1350,567)(1330,567)
\thicklines \path(198,646)(218,646)
\thicklines \path(1350,646)(1330,646)
\put(176,646){\makebox(0,0)[r]{$0.20$}}
\thicklines \path(198,726)(218,726)
\thicklines \path(1350,726)(1330,726)
\thicklines \path(198,805)(218,805)
\thicklines \path(1350,805)(1330,805)
\put(176,805){\makebox(0,0)[r]{$0.24$}}
\thicklines \path(198,885)(218,885)
\thicklines \path(1350,885)(1330,885)
\thicklines \path(198,90)(198,110)
\thicklines \path(198,885)(198,865)
\put(198,45){\makebox(0,0){$0.0$}}
\thicklines \path(428,90)(428,110)
\thicklines \path(428,885)(428,865)
\put(428,45){\makebox(0,0){$0.2$}}
\thicklines \path(659,90)(659,110)
\thicklines \path(659,885)(659,865)
\put(659,45){\makebox(0,0){$0.4$}}
\put(774,-50){\makebox(0,0){$\theta/(2\pi)$}}
\thicklines \path(889,90)(889,110)
\thicklines \path(889,885)(889,865)
\put(889,45){\makebox(0,0){$0.6$}}
\thicklines \path(1120,90)(1120,110)
\thicklines \path(1120,885)(1120,865)
\put(1120,45){\makebox(0,0){$0.8$}}
\thicklines \path(1350,90)(1350,110)
\thicklines \path(1350,885)(1350,865)
\put(1350,45){\makebox(0,0){$1.0$}}
\thicklines \path(198,90)(1350,90)(1350,885)(198,885)(198,90)
\thinlines \path(198,101)(198,101)(199,100)(201,100)(202,100)(204,100)(205,100)(206,100)(208,100)(209,100)(210,100)(212,100)(213,100)(215,100)(218,100)(221,101)(227,103)(232,105)(245,112)(258,122)(269,133)(296,166)(344,246)(391,336)(440,434)(488,523)(534,599)(583,667)(630,720)(654,742)(680,763)(703,778)(728,791)(752,799)(765,803)(771,804)(778,805)(784,806)(789,807)(793,807)(796,807)(797,807)(799,807)(800,807)(801,807)(802,807)(803,807)(805,807)(806,807)(808,807)(809,807)
\thinlines \path(809,807)(810,807)(813,807)(816,807)(818,807)(824,806)(830,805)(837,804)(848,802)(860,798)(873,793)(897,781)(922,764)(972,717)(1020,654)(1067,579)(1116,483)(1164,382)(1209,284)(1258,191)(1281,157)(1293,141)(1305,128)(1316,118)(1326,110)(1332,107)(1338,104)(1341,103)(1344,102)(1350,101)
\end{picture}
\end{minipage}
\end{center}
\caption{Partially integrated branching ratio
${\cal B} \equiv \int^8_1ds\frac{d{\cal B}}{ds}$ and FB asymmetry 
$\bar{{\cal
A}}\equiv\frac{\int^8_1dsd{\cal A}/ds}{\int^8_1dsd{\cal B}/ds}$. 
The angle $\theta$ is defined by $\frac{C_{SL}^N}{C_{BL}} \equiv
\tan\theta$. The coefficients $C_{SL}^N$ and $C_{BR}$ move under the
conditions (2.7).}\label{fig:partialbr}
\end{figure}
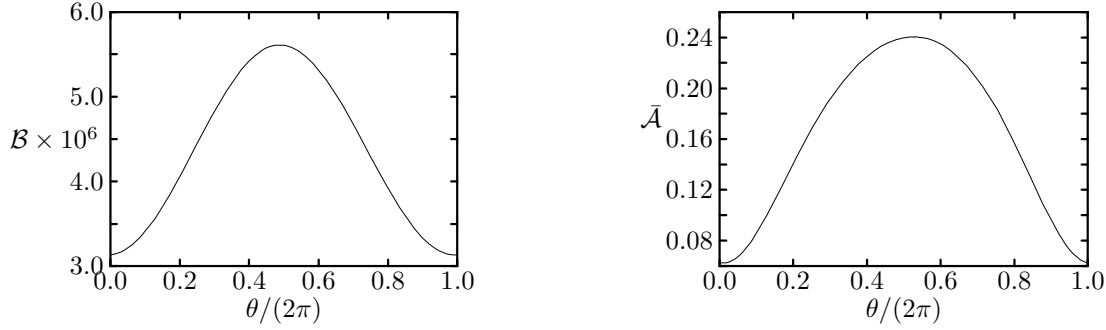

We show the branching ratio of $B\rightarrow X_s l^+ l^- $ for
massless lepton case in Figure
\ref{fig:brratio} in the absence of any new local interactions, 
but with new non-local interactions and the already existing  local operators
of the SM.   In this case, the branching ratio is given as
\begin{eqnarray}
	\frac{d{\cal B}}{ds} (B\rightarrow X_s l^+ l^- ) 
	&=& \frac{1}{2m_b^8}{\cal B}_0
		[S_1(s)m_b^2\{\left|C_{SL}^N\right|^2 +
				\left|C_{BR}\right|^2\}
                + 2 S_2(s)m_b^2 Re[C_{SL}^N C_{BR}^*] \nonumber\\
		&&+4 S_3(s)m_bm_sRe[C_{SL}^NC_9^{eff*}]
				+4 S_4(s)m_b^2Re[C_{BR}C_9^{eff*}]\nonumber\\
		&&+M_2(s)\{\left|C_9^{eff}-C_{10}\right|^2 +
	\left|C_9^{eff}+C_{10}\right|^2\}], \label{eqn:newnonlocal}
\end{eqnarray}
with $s = (p_{l^+} + p_{l^-})^2$, invariant mass-square of lepton pair.
(The case with the all twelve operators is given in Apendix.)
The $S_n(s)$ and $M_2(s)$ are given in Refs. \cite{FKMY,FKY}. 
We note that the branching ratio is more sensitive to the change of $C_{BR}$
than of $C_{SL}$, as shown in Figure \ref{fig:brratio}, 
because the interferences of the $C_{SL}$ and the $C_{BR}$ to the vector
type interactions of the SM give
\begin{eqnarray}
Tr\{\bar{s}\sigma_{\mu\nu}Lb (\bar{s}_L\gamma_{\rho}b_L)^*\}
		&\propto& m_s, \nonumber \\
{\rm and}~~~
Tr\{\bar{s}\sigma_{\mu\nu}Rb (\bar{s}_L\gamma_{\rho}b_L)^*\}
		&\propto& m_b, \nonumber
\end{eqnarray}
respectively. 
The contribution from these coefficients
oscillates as we vary the values of $(C_{SL}^N,~C_{BR})$ within the 
region $II$ in Figure \ref{fig:brslconstraint}, 
because of the constraint~(\ref{eqn:slbrconstraint}). 
However, if we account  only for the $C_{BR}$ (as in the SM),
or equivalently, if we assume $m_s = 0$, then $C_{BR}$ moves only between
$-2\left|C_7^{eff}\right|$ and $2\left|C_7^{eff}\right|$, 
and the branching ratio decreases monotonously.
The points $(C_{SL}^N, C_{BR}) =(0,2\left|C_7^{eff}\right|)$ and 
$(0,-2\left|C_7^{eff}\right|)$ are the minimum and the maximum values. 
(The SM value corresponds almost to  the point $(0, -2 C_7^{eff})$.)
This behavior reappears for the partially integrated branching ratio
${\cal B}\equiv\int^8_1ds\frac{d{\cal B}}{ds}$ \cite{FKMY} as shown  
in Figure \ref{fig:partialbr}.

Now we consider the forward-backward (FB) asymmetry defined as
\begin{equation}
	\frac{d\bar{\cal A}}{ds}
       \equiv \frac{d{\cal A}/ds}{d{\cal B}/ds}
        = \frac{
                 \int^1_0dz\frac{d^2{\cal B}}{dsdz}
                -\int^0_{-1}dz\frac{d^2{\cal B}}{dsdz}}
             {\int^1_0dz\frac{d^2{\cal B}}{dsdz}
                +\int^0_{-1}dz\frac{d^2{\cal B}}{dsdz}},
			\label{eqn:fbasymmetry}
\end{equation}
where $z$ is the cosine value of the angle between the momentum of $B$ meson 
and that of $l^+$ in the laboratory frame.
We also show the normalized FB asymmetry curves, $d\bar{{\cal A}}/ds$, 
at the four points 
$
(C_{SL}^N, C_{BR})=(0, -2C_7^{eff}),~~(0, 2C_7^{eff}),~~(-2C_7^{eff}, 0),~~(2C_7^{eff}, 0)
$ 
in Figure \ref{fig:brratio}.
The line for the $(0, -2C_7^{eff})$ corresponds to the SM result.  
The unnormalized FB asymmetry of $B\rightarrow X_s l^+ l^- $ for
massless lepton case in the absence of any new local interactions, 
but with new non-local interactions and the already existing local operators of
the SM is given as  
\begin{eqnarray}
	\frac{d{\cal A}}{ds}&=&\frac{1}{2m_b^8}{\cal B}_0 u(s)^2[
		8(Re\{(m_b^2C_{BR}+m_bm_sC_{SL}^N)C_{10}^*)\nonumber\\
		&&+2s(\left|C_9^{eff}-C_{10}\right|^2
		-\left|C_9^{eff}+C_{10}\right|^2)].\label{eqn:FBnonlocal}
\end{eqnarray}
(The case for massive lepton with the all 12 operators is given in
Appendix.)
We note that, as is the case for the branching ratio, 
the FB asymmetry is more sensitive to the change of $C_{BR}$
than  of $C_{SL}$ as shown in Figure \ref{fig:brratio}, and
the oscillating behavior reappears. To see
the sensitivity of the asymmetry for each coefficient, we introduce the
partially integrated (un)normalized FB asymmetry $\bar{{\cal A}}$ (${\cal A}$)
defined as
\begin{eqnarray*}
	\bar{\cal A}&\equiv&\frac{{\cal A}}{{\cal B}},\\
	{\cal A}&\equiv&\int^8_1ds\frac{d{\cal A}}{ds},
\end{eqnarray*}
where ${\cal B}\equiv\int^8_1ds\frac{d{\cal B}}{ds}$. We present the
influence of two non-local coefficients on the normalized FB asymmetry 
in Figure \ref{fig:partialbr}. 

\setcounter{equation}{0}
\section{Discussions and Conclusions}\label{sec:behavior}
\begin{figure}
\begin{center}
\setlength{\unitlength}{0.180900pt}
\begin{picture}(1500,900)(0,0)
\footnotesize
\put(800,0){\makebox(0,0){${\cal B}\times 10^6$}}
\put(50,400){\makebox(0,0)[r]{$\bar{{\cal A}}$}}
\put(599,520){\raisebox{-1.2pt}{\makebox(0,0){$\Diamond$}}}
\put(880,734){\makebox(0,0){$+$}}
\put(748,645){\raisebox{-1.2pt}{\makebox(0,0){$\Box$}}}
\put(731,667){\makebox(0,0){$\times$}}
\thicklines \path(220,90)(240,90)
\thicklines \path(1394,90)(1374,90)
\put(198,90){\makebox(0,0)[r]{$-0.25$}}
\thicklines \path(220,156)(240,156)
\thicklines \path(1394,156)(1374,156)
\put(198,156){\makebox(0,0)[r]{$-0.20$}}
\thicklines \path(220,222)(240,222)
\thicklines \path(1394,222)(1374,222)
\put(198,222){\makebox(0,0)[r]{$-0.15$}}
\thicklines \path(220,288)(240,288)
\thicklines \path(1394,288)(1374,288)
\put(198,288){\makebox(0,0)[r]{$-0.10$}}
\thicklines \path(220,354)(240,354)
\thicklines \path(1394,354)(1374,354)
\put(198,354){\makebox(0,0)[r]{$-0.05$}}
\thicklines \path(220,420)(240,420)
\thicklines \path(1394,420)(1374,420)
\put(198,420){\makebox(0,0)[r]{$0.00$}}
\thicklines \path(220,486)(240,486)
\thicklines \path(1394,486)(1374,486)
\put(198,486){\makebox(0,0)[r]{$0.05$}}
\thicklines \path(220,552)(240,552)
\thicklines \path(1394,552)(1374,552)
\put(198,552){\makebox(0,0)[r]{$0.10$}}
\thicklines \path(220,618)(240,618)
\thicklines \path(1394,618)(1374,618)
\put(198,618){\makebox(0,0)[r]{$0.15$}}
\thicklines \path(220,684)(240,684)
\thicklines \path(1394,684)(1374,684)
\put(198,684){\makebox(0,0)[r]{$0.20$}}
\thicklines \path(220,750)(240,750)
\thicklines \path(1394,750)(1374,750)
\put(198,750){\makebox(0,0)[r]{$0.25$}}
\thicklines \path(220,90)(220,110)
\thicklines \path(220,750)(220,730)
\put(220,45){\makebox(0,0){$0$}}
\thicklines \path(455,90)(455,110)
\thicklines \path(455,750)(455,730)
\put(455,45){\makebox(0,0){$2$}}
\thicklines \path(690,90)(690,110)
\thicklines \path(690,750)(690,730)
\put(690,45){\makebox(0,0){$4$}}
\thicklines \path(924,90)(924,110)
\thicklines \path(924,750)(924,730)
\put(924,45){\makebox(0,0){$6$}}
\thicklines \path(1159,90)(1159,110)
\thicklines \path(1159,750)(1159,730)
\put(1159,45){\makebox(0,0){$8$}}
\thicklines \path(1394,90)(1394,110)
\thicklines \path(1394,750)(1394,730)
\put(1394,45){\makebox(0,0){$10$}}
\thicklines \path(220,90)(1394,90)(1394,750)(220,750)(220,90)
\put(807,818){\makebox(0,0){$$}}
%
\thinlines \path(1394,227)(1224,244)(1137,257)(1049,272)(968,290)(900,309)(784,354)(737,380)(694,409)(657,441)(631,469)(612,496)(600,518)(596,527)(595,531)(595,533)(594,534)(594,535)(594,536)(594,537)(594,537)(594,538)(594,539)(594,539)(594,540)(594,541)(594,541)(594,542)(595,542)(595,543)(595,544)(596,545)(597,546)(597,546)(598,547)(599,547)(600,548)(601,548)(601,548)(601,548)(602,548)(602,548)(603,548)(603,548)(603,548)(604,548)(604,548)(605,548)(605,548)(606,548)(607,548)
\thinlines \path(607,548)(609,547)(612,547)(617,545)(622,543)(674,519)(759,480)(870,440)(1003,404)(1173,371)(1365,344)(1394,341)
\thicklines \dashline[-10]{10}(1394,274)(1346,278)(1246,290)(1145,304)(981,335)(911,354)(844,375)(744,417)(665,463)(618,500)(605,513)(601,518)(599,519)(599,520)(598,521)(598,521)(598,521)(598,521)(598,521)(598,521)(598,521)(598,521)(598,521)(598,521)(598,521)(598,521)(598,521)(599,521)(599,521)(600,520)(608,514)(649,483)(718,441)(811,399)(869,378)(939,358)(1012,341)(1101,324)(1287,297)(1390,285)(1394,285)
\thinlines \dashline[-10]{5}(1394,642)(1284,637)(1098,626)(1017,619)(939,611)(819,594)(767,584)(719,573)(682,561)(648,549)(624,538)(609,530)(604,526)(601,523)(600,522)(599,522)(599,521)(599,521)(599,521)(599,520)(599,520)(599,520)(599,520)(599,520)(599,520)(599,519)(599,519)(599,519)(599,519)(600,519)(600,519)(600,518)(601,518)(601,518)(601,518)(602,518)(602,518)(602,518)(603,518)(603,518)(603,518)(604,518)(604,518)(604,518)(605,518)(605,518)(607,518)(608,518)(611,519)(617,520)(667,531)
\thinlines \dashline[-10]{5}(667,531)(743,547)(844,564)(980,581)(1139,594)(1339,606)(1394,609)
\Thicklines \path(598,520)(598,520)(598,520)(599,520)(599,520)(599,520)(599,520)(599,520)(599,520)(599,520)(599,520)(599,520)(599,520)(600,520)(600,520)(600,520)(602,521)(605,523)(622,539)(647,563)(677,591)(714,621)(752,648)(787,672)(821,693)(848,709)(867,721)(874,726)(879,730)(880,731)(880,731)(880,732)(880,732)(880,732)(881,732)(881,732)(881,733)(881,733)(881,733)(881,733)(881,733)(881,733)(881,733)(881,733)(880,733)(880,734)(880,734)(880,734)(880,734)(880,734)(880,734)
\Thicklines \path(880,734)(879,734)(879,734)(879,735)(878,735)(878,735)(878,735)(878,735)(878,735)(877,735)(877,735)(877,735)(877,735)(876,735)(876,735)(876,735)(876,735)(875,735)(874,735)(874,735)(872,734)(869,734)(865,733)(855,730)(829,721)(799,707)(762,688)(726,663)(691,636)(658,604)(631,573)(612,547)(605,536)(601,528)(599,525)(599,524)(599,523)(598,522)(598,522)(598,522)(598,522)(598,521)(598,521)(598,521)(598,521)(598,521)(598,521)(598,521)(598,520)(598,520)(598,520)
\Thicklines \path(1394,733)(1093,739)(847,746)(741,750)
\Thicklines \path(367,750)(356,746)(346,742)(335,736)(324,727)(314,717)(305,703)(297,688)(290,668)(283,644)(277,616)(273,587)(269,551)(265,507)(261,455)(259,409)(258,383)(258,358)(257,347)(257,334)(257,324)(257,313)(257,307)(257,301)(257,295)(257,290)(257,280)(257,275)(257,270)(257,261)(257,253)(257,247)(257,240)(258,233)(258,227)(258,222)(259,217)(259,213)(259,210)(260,207)(260,205)(261,203)(261,202)(262,201)(262,201)(263,200)(264,200)(264,201)(265,201)(266,202)(267,206)
\Thicklines \path(267,206)(269,210)(289,269)(304,306)(322,344)(345,379)(368,407)(393,430)(424,452)(487,485)(527,500)(567,512)(650,531)(752,548)(859,561)(986,572)(1127,581)(1271,589)(1394,594)
\end{picture}
\end{center}
\caption{
Correlation between the partially integrated branching ratio ${\cal B}$ 
and the partially integrated normalized forward-backward asymmetry
$\bar{{\cal A}}$ for the non-local interactions (thick solid ellipse) and the
vector-type interactions, $C_{LL}$ (another thick solid line), 
$C_{LR}$ (thin solid line), $C_{RL}$ (thick dotted line) 
and $C_{RR}$ (thin dotted line). The
marks $\Diamond$, $\Box$, $+$ and $\times$ correspond to 
$(C_{SL}^N, C_{BR}) = (0, -2C_7^{eff})$, $(-2C_7^{eff}, 0)$, 
$(0, 2C_7^{eff})$ and $(2C_7^{eff}, 0)$, respectively. The point
for the SM is placed just near the $(0, -2 C_7^{eff})$.}\label{fig:correlation}
\end{figure}
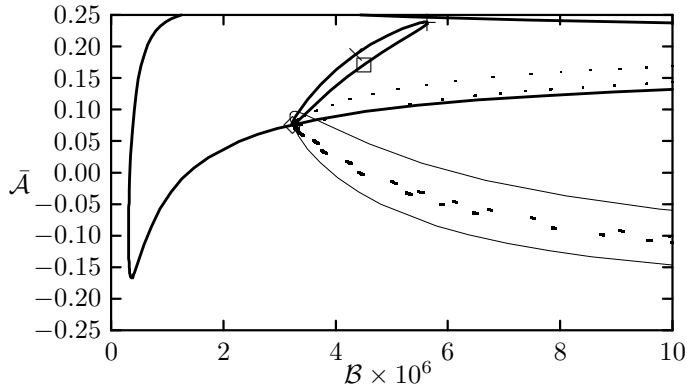

In Sec.\ref{sec:decaydistrib}, we investigated both the branching ratio and the FB asymmetry
independently.  
As shown,  both observables are more sensitive to changes of
$C_{BR}$ than to those of $C_{SL}$, 
and oscillate as the non-local coefficients
change.  Now we show the correlation between the branching ratio
and the FB asymmetry in Figure \ref{fig:correlation}. 
It is very interesting to compare
the correlation for various interactions, because the flows in the
plane $({\cal B}, \bar{{\cal A}})$ depend on interactions which we consider.
We already investigated the correlation flows for the case of local
interactions in \cite{FKMY}. 
The flows in the plain $({\cal B}, \bar{{\cal A}})$ 
for the non-local interactions are quite different from the local ones. 
As found in Ref. \cite{FKMY}, 
the standard model point is just near 
$(C_{SL}^N, C_{BR})=(0, 2\left|C_7^{eff}\right|)$ in the plane,
so that it is placed at the lowest point (marked as $\Diamond$) 
of the closed ellipse in Figure \ref{fig:correlation}. 
Therefore, if there exist any non-local interactions in new theory
beyond standard model, both  the ratio and the FB asymmetry monotonically
increase. 

However, the vector-type
interactions increase or decrease the branching ratio (the FB asymmetry) 
as the values of  vector-type coefficients increase (decrease) or decrease
(increase).  And the scalar-type and tensor-type interactions make no
change for the unnormalized FB asymmetry. 
We can also understand the presented behavior for
the non-local interactions with the following arguments: 
The branching ratio and the FB asymmetry change, 
for $C_{BR}$ and $C_{SL}$, only through the second
term in Eq. (\ref{eqn:newnonlocal}) and the first term in
Eq. (\ref{eqn:FBnonlocal}), 
since the two coefficients cannot change simultaneously under the
condition (\ref{eqn:slbrconstraint}). 
If we leave the leading term in $m_s$, 
the partially integrated branching ratio (${\cal B}$) and the partially 
integrated forward-backward asymmetry (${\cal A}$) are expressed as
\begin{eqnarray*}
	{\cal B} &=& {\cal B}_{c1} + {\cal B}_{c2} [m_b C_{BR} (9-2m_b^2)
		+m_s C_{SL}^N (9+2m_b^2)] + {\cal O}(m_s^2),  ~~({\cal B}_{c2} > 0),\\
	{\cal A} &=&
	{\cal A}_{c1} - {\cal A}_{c2} (m_b C_{BR} + m_s C_{SL}^N) + {\cal O}(m_s^2),
		~~({\cal A}_{c2} > 0),
\end{eqnarray*}
where ${\cal B}_{c1}$, ${\cal B}_{c2}$, ${\cal A}_{c1}$ 
and ${\cal A}_{c2}$ are
independent of $C_{BR}$ and $C_{SL}$ and $m_s$. Under the condition
$C_{SL}=\sqrt{4 \left(C_7^{eff}\right)^2 - C_{BR}^2}$, 
${\cal B}$ and ${\cal A}$ achieve the minimum and the maximum at
$C_{BR} = 2 \left|C_7^{eff}\right|$ and $-2\left|C_7^{eff}\right|$. 
Therefore we will be able to know the sign of the $C_7^{eff}$ 
and deviation from the predictions of the SM by using this correlation.

\begin{figure}
\begin{center}
\setlength{\unitlength}{0.180900pt}
\begin{picture}(1500,900)(0,0)
\footnotesize
\thicklines \path(198,90)(218,90)
\thicklines \path(1394,90)(1374,90)
\put(176,90){\makebox(0,0)[r]{$-0.3$}}
\thicklines \path(198,215)(218,215)
\thicklines \path(1394,215)(1374,215)
\put(176,215){\makebox(0,0)[r]{$-0.2$}}
\thicklines \path(198,340)(218,340)
\thicklines \path(1394,340)(1374,340)
\put(176,340){\makebox(0,0)[r]{$-0.1$}}
\thicklines \path(198,465)(218,465)
\thicklines \path(1394,465)(1374,465)
\put(176,465){\makebox(0,0)[r]{$0$}}
\put(50,465){\makebox(0,0)[r]{$\bar{{\cal A}}$}}
\thicklines \path(198,591)(218,591)
\thicklines \path(1394,591)(1374,591)
\put(176,591){\makebox(0,0)[r]{$0.1$}}
\thicklines \path(198,716)(218,716)
\thicklines \path(1394,716)(1374,716)
\put(176,716){\makebox(0,0)[r]{$0.2$}}
\thicklines \path(198,841)(218,841)
\thicklines \path(1394,841)(1374,841)
\put(176,841){\makebox(0,0)[r]{$0.3$}}
\thicklines \path(198,90)(198,110)
\thicklines \path(198,841)(198,821)
\put(198,45){\makebox(0,0){$0$}}
\thicklines \path(437,90)(437,110)
\thicklines \path(437,841)(437,821)
\put(437,45){\makebox(0,0){$2$}}
\thicklines \path(676,90)(676,110)
\thicklines \path(676,841)(676,821)
\put(676,45){\makebox(0,0){$4$}}
\put(800,10){\makebox(0,0){${\cal B}\times 10^6$}}
\thicklines \path(916,90)(916,110)
\thicklines \path(916,841)(916,821)
\put(916,45){\makebox(0,0){$6$}}
\thicklines \path(1155,90)(1155,110)
\thicklines \path(1155,841)(1155,821)
\put(1155,45){\makebox(0,0){$8$}}
\thicklines \path(1394,90)(1394,110)
\thicklines \path(1394,841)(1394,821)
\put(1394,45){\makebox(0,0){$10$}}
\thicklines \path(198,90)(1394,90)(1394,841)(198,841)(198,90)
\Thicklines \path(1394,761)(1146,765)(905,769)(786,772)(689,774)(644,775)(621,775)(598,776)(588,776)(578,776)(572,776)(567,776)(562,776)(557,776)(552,776)(548,776)(543,776)(538,776)(534,776)(529,776)(525,776)(521,776)(512,776)(504,776)(494,775)(485,775)(470,775)(454,774)(437,773)(420,771)(406,769)(391,766)(378,763)(364,760)(353,755)(342,750)(330,743)(319,734)(308,724)(300,712)(291,699)(283,682)(277,665)(271,643)(264,616)(259,582)(254,546)(250,509)(247,470)(244,426)(242,384)
\Thicklines \path(242,384)(241,346)(240,324)(240,314)(240,305)(240,296)(240,291)(240,286)(240,281)(240,276)(240,272)(240,267)(240,260)(240,256)(240,253)(240,246)(240,239)(240,234)(241,229)(241,225)(241,221)(242,218)(242,215)(243,212)(243,211)(243,210)(244,209)(244,208)(245,208)(245,207)(246,207)(246,208)(247,209)(248,210)(248,211)(252,219)(274,288)(305,360)(326,395)(348,423)(398,468)(430,489)(462,506)(499,521)(537,534)(614,554)(706,571)(801,585)(906,596)(1032,606)(1161,614)
\Thicklines \path(1161,614)(1310,622)(1394,625)
\Thicklines \dashline[-20]{1}(1394,612)(1315,608)(1159,598)(1026,588)(914,577)(803,563)(695,544)(650,533)(604,521)(564,509)(524,493)(458,461)(426,439)(399,416)(370,386)(346,354)(308,282)(292,239)(277,192)(265,149)(261,131)(258,123)(256,117)(254,112)(254,111)(253,110)(253,109)(252,109)(252,109)(251,109)(251,109)(251,109)(250,110)(250,111)(249,113)(249,115)(249,117)(248,120)(248,124)(248,128)(247,133)(247,138)(247,143)(247,149)(247,156)(247,159)(247,163)(247,171)(247,180)(247,188)(247,196)
\Thicklines \dashline[-20]{1}(247,196)(247,204)(248,221)(249,260)(251,306)(253,350)(257,397)(261,444)(270,517)(276,552)(283,581)(289,604)(297,627)(312,661)(321,676)(330,689)(342,701)(353,711)(378,727)(392,734)(408,740)(437,748)(454,751)(470,754)(486,756)(503,758)(536,761)(556,762)(574,763)(595,764)(618,765)(630,765)(643,765)(655,766)(668,766)(680,766)(693,766)(705,766)(716,766)(722,766)(728,766)(734,766)(739,766)(746,766)(749,766)(753,766)(759,766)(765,766)(772,766)(779,766)(791,766)
\Thicklines \dashline[-20]{1}(791,766)(806,766)(820,766)(873,766)(925,765)(1169,762)(1394,760)
\thinlines \path(1394,701)(1325,700)(1221,699)(1129,698)(969,695)(889,694)(821,692)(755,689)(688,686)(629,683)(570,678)(521,673)(478,667)(435,658)(400,649)(368,636)(351,628)(337,619)(323,608)(312,597)(293,572)(284,557)(276,541)(265,511)(258,482)(256,473)(255,469)(255,466)(255,465)(255,465)(255,465)(255,465)(255,464)(255,464)(255,464)(255,464)(255,464)(255,464)(255,464)(255,464)(255,464)(255,465)(256,466)(261,482)(271,509)(284,538)(301,563)(312,576)(323,587)(348,606)(364,615)
\thinlines \path(364,615)(381,623)(417,636)(455,646)(498,654)(549,661)(602,667)(663,672)(725,675)(786,678)(929,684)(1014,686)(1099,688)(1274,691)(1394,692)
\thinlines \dashline[-10]{5}(1394,685)(1351,685)(1254,683)(1152,681)(984,677)(907,674)(829,671)(760,668)(690,663)(628,658)(575,653)(526,646)(478,638)(437,628)(397,615)(379,608)(362,599)(346,589)(333,580)(309,557)(298,543)(288,529)(263,476)(259,467)(256,460)(256,459)(255,459)(255,458)(255,458)(255,458)(255,458)(255,458)(254,458)(254,458)(254,458)(254,459)(254,459)(254,459)(254,459)(254,460)(254,460)(254,460)(254,461)(254,461)(254,462)(254,462)(254,463)(254,465)(255,470)(256,481)(259,496)
\thinlines \dashline[-10]{5}(259,496)(263,514)(275,550)(283,569)(291,585)(311,611)(324,624)(337,634)(364,649)(382,656)(399,662)(419,668)(438,672)(478,679)(525,685)(572,689)(624,693)(685,696)(748,698)(819,700)(893,702)(975,703)(1137,705)(1226,706)(1326,707)(1394,707)
\Thicklines \path(573,544)(573,544)(573,544)(573,544)(573,544)(573,544)(573,544)(573,544)(573,544)(573,544)(573,544)(574,544)(574,544)(574,544)(574,544)(575,544)(577,545)(579,547)(597,564)(623,590)(655,619)(694,650)(734,678)(771,702)(807,724)(835,741)(855,753)(863,758)(867,761)(868,763)(868,763)(869,764)(869,764)(869,764)(869,764)(869,764)(869,764)(869,765)(869,765)(869,765)(869,765)(869,765)(869,765)(869,765)(869,765)(869,765)(869,766)(869,766)(868,766)(868,766)(868,766)
\Thicklines \path(868,766)(868,766)(867,766)(867,766)(867,766)(866,766)(866,766)(866,766)(866,766)(866,767)(865,767)(865,767)(865,767)(864,767)(864,767)(864,767)(864,766)(863,766)(862,766)(862,766)(860,766)(857,766)(852,765)(842,762)(815,753)(783,739)(745,718)(706,693)(670,665)(635,632)(607,600)(587,572)(580,560)(575,552)(574,549)(573,548)(573,546)(573,546)(573,546)(573,545)(573,545)(573,545)(573,545)(573,545)(573,545)(573,544)(573,544)(573,544)(573,544)(573,544)(573,544)
\put(573,544){\raisebox{-1.2pt}{\makebox(0,0){$\Diamond$}}}
\put(869,766){\makebox(0,0){$+$}}
\put(730,676){\raisebox{-1.2pt}{\makebox(0,0){$\Box$}}}
\put(712,697){\makebox(0,0){$\times$}}
\end{picture}
\end{center}
\caption{Correlation as the Wilson coefficients $C_{LL}$ moves, receiving
the effects of the interference between the new vector-type interaction 
and the new non-local interactions, whose Wilson coefficients are 
$(C_{SL}^N,C_{BR}) = (0, -2 C_7^{eff})$ (thick solid line), 
$(0, 2 C_7^{eff})$ (thick dashed line), $(-2 C_7^{eff}, 0)$
(thin solid line) and $(2 C_7^{eff}, 0)$ (thin dashed line). 
To refer, we also show
the correlation as the set of the coefficients moves. The latter is also
described in Figure 6 with the same notation.}\label{fig:CLLnonlocal}
\end{figure}
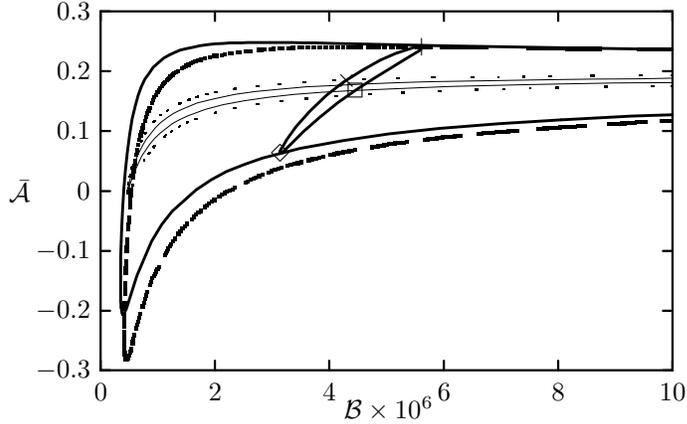

The non-local interactions are dominant as the momentum transfer 
from the $b$-quark to the $s$-quark gets comparable with the lepton mass, 
that is, as $\sqrt{s} \to m_l \sim 0$, 
because of the factor $1/s$. But, the sensitivity of the ratio to
the changes of $C_{BR}$ and $C_{SL}$ is not so large in comparison with the
local interactions, because of the  constraint
(\ref{eqn:slbrconstraint}).
If there is a new local interaction in addition, 
understanding the interference between the non-local interactions 
and the new local interaction would be extremely important 
within the small and
non-vanishing $s$ region, $1 < s < 8$ ~GeV$^2$ \cite{Lu}. 
For example, in the massless
limit, the scalar- and tensor-type interactions cannot contribute to the FB
asymmetry. Hence, if we find a deviation of the FB asymmetry from the SM
prediction,  we can infer that there are new non-scalar- or non-tensor-type
interactions. To extend further our discussion, 
suppose that interactions which act on massless leptons 
are equal to the ones which act on massive leptons. 
If we cannot find the lepton longitudinal polarization asymmetry
$\left<P_L^+\right> + \left<P_L^-\right>$ from the precise experiments for the
decays $B\to X_s\tau^+\tau^-$, then we can conclude there is no scalar- or
tensor-type interaction \cite{FKY}, and infer that there are 
vector-type interactions like the SM and non-local interactions as well. 
In such a case, we should consider the
interference between the non-local interactions and the vector-type ones. In
Figure \ref{fig:CLLnonlocal}, we show the flow as $C_{LL}$ moves when there
are new non-local interactions, where, again, the branching ratio and the
FB asymmetry are integrated over $s$ from 1 GeV$^2$ to 8 GeV$^2$. 


To summarize, we investigated the
effects of the non-local interactions in the rare $B$ decays $\Bsll$ in the
model-independent way. 
In our model-independent analysis in this paper and Refs. \cite{FKMY,FKY}, we
used all the operators which influence  the process $\Bsll$, those are, ten
local and two non-local four-Fermi operators. 
We, here, studied the sensitivity to the coefficients of the
non-local interactions for the branching ratio and the forward-backward (FB)
asymmetry. We note that both  the ratio and the FB asymmetry are
more sensitive to $C_{BR}$
than to $C_{SL}$. We did not use the $C_{SL}$
introduced at first in Eq. (\ref{eqn:matrix}), instead we used the normalized
Wilson coefficient $C_{SL}^N \equiv \frac{m_s}{m_b} C_{SL}$, in order not to
mislead. Nevertheless, the interference terms between the $C_{SL}^N$ 
and the other local
operators include an extra  mass ratio $m_s/m_b$, compared 
to the interferences from the $C_{BR}$ and others, 
and therefore, the operator $C_{BR}$ gives greater influence on the
ratio and  the FB asymmetry than the $C_{SL}$. 
Consequently, the value of the $C_{BR}$ almost decides the size of the
branching ratio and the FB asymmetry as the result of their correlation.
If there is any new charged local interaction, which contributes to
$\Bsll$, like in the minimal supersymmetric standard model, 
we must consider appropriate non-local interactions, 
because any charged particle's interaction with
photons yields non-local interactions. 
Especially, in the small invariant mass region,
we cannot ignore the contribution from the non-local interactions. 
And our analysis would give very useful help for the precise study of
new physics in $\Bsll$ when such a new local interaction exists. 
\\
\bigskip\bigskip

%
%
\centerline{\Large \bf ACKNOWLEDGMENTS}

\noindent
We would like to G. Cvetic and T.Morozumi for very useful suggestions and
comments. The work of C.S.K. was supported 
in part by  BK21 Project, SRC Program and Grant No. 2000-1-11100-003-1
of the KOSEF, and in part by the KRF Grants (Project No.
1997-011-D00015 and Project No. 2000-015-DP0077).
The work of T.Y. was supported in part by Grant-in-Aid for
Scientific Research from the Ministry of Education, Science and
Culture of Japan and in part by JSPS Research Fellowships for Young
Scientists.\\

\newpage

\appendix

\setcounter{equation}{0}
\renewcommand{\theequation}{\Alph{section}.\arabic{equation}}
\newcommand{\mymyequation}[1]{\equation{#1}\setcounter{equation}{1}}
\centerline{\Large \bf APPENDIX}
\section{Branching Ratio and the Forward-backward Asymmetry with complete 12
Operators}\label{apen:kinfun}

In Ref. \cite{FKMY}, we have already studied the
differential branching ratio  and the FB asymmetry for massless leptons
without including two non-local interactions, and in Ref. \cite{FKY} 
we investigated the differential branching ratio and the
polarization asymmetries for massive leptons with including all 12 operators. 
Here we show the differential branching ratio for massless leptons 
by including two new non-local interactions,
\begin{eqnarray}
	\frac{d{\cal B}(s)}{ds} = \frac{1}{2m_b^8}{\cal B}_0&[&S_1(s)
		\{m_s^2|C_{SL}|^2 + m_b^2|C_{BR}|^2 \}\nonumber\\
		&&+S_2(s)\{2m_bm_sRe[C_{SL}C_{BR}^*]\}\nonumber\\
		&&+S_3(s)\{2m_s^2Re[C_{SL}(C_{LL}^*+C_{LR}^*)]
		+2m_bm_sRe[C_{BR}(C_{RL}^*+C_{RR}^*)]\}\nonumber\\
		&&+S_4(s)\{2m_b^2Re[C_{BR}(C_{LL}^*+C_{LR}^*)]
		+2m_bm_sRe[C_{SL}(C_{RL}^*+C_{RR}^*)]\}\nonumber\\
		&&+S_5(s)\{2(m_sC_{SL} + m_bC_{BR})C_T^*\}\nonumber\\
		&&+S_6(s)\{4(m_bC_{BR} - m_sC_{SL})C_{TE}^*\}\nonumber\\
		&&+M_2(s) \{|C_{LL}|^2 + |C_{LR}|^2+|C_{RL}|^2+|C_{RR}|^2\}\nonumber\\
		&&-M_6(s)\{2Re[C_{LL}C_{RL}^* + C_{LR}C_{RR}^*]\nonumber\\
		&&~~~~~~~~~-Re[C_{LRLR}C_{RLLR}^* + C_{LRRL}C_{RLRL}^*]\}\nonumber\\
		&&+M_8(s)\{|C_{LRLR}|^2+|C_{RLLR}|^2+|C_{LRRL}|^2+|C_{RLRL}|^2\}\nonumber\\
        &&+M_9(s)\{16|C_T|^2+64|C_{TE}|^2\}
		.\label{eqn:brratio}
\end{eqnarray}
The kinematic functions, $S_n(s)$ and $M_n(s)$, are all given in
detail in Ref. \cite{FKMY,FKY}.
We find from Eq. (\ref{eqn:brratio}) that $S_2(s)$, which includes $m_s$ as an
overall factor, is multiplied by $m_b$ and $m_s$ and, therefore, it is 
negligible. 
We also show the most general form of the FB asymmetry
in case of massive leptons, which includes all 12 operators. 
It is as follows:
\begin{eqnarray}
	\frac{d{\cal A}}{ds}&=&\frac{1}{2m_b^8}{\cal B}_0u(s)^2[
		-4(Re\{m_b^2C_{BR}+m_s^2C_{SL})(C_{LL}^*-C_{LR}^*\})\nonumber\\
	&&+8m_bm_sRe\{(C_{BR}+C_{SL})(C_{RL}^*-C_{RR}^*)\}\nonumber\\
	&&+4m_sm_lRe\{C_{SL}(C_{RLLR}^*+C_{RLRL}^*)\}\nonumber\\
	&&+4m_bm_lRe\{C_{BR}(C_{LRLR}^*+C_{LRRL}^*)\}\nonumber\\
	&&+2s(\left|C_{LL}\right|^2 - \left|C_{LR}\right|^2 
		- \left|C_{RL}\right|^2 + \left|C_{RR}\right|^2)\nonumber\\
	&&-8s(Re\{C_{LRLR}(C_T^*-2C_{TE}^*)\}+Re\{C_{RLRL}(C_T^*+2C_{TE}^*)\})\nonumber\\
	&&-2m_bm_l(Re\{(C_{LL}+C_{LR})(C_{LRLR}^*+C_{LRRL}^*)\}\nonumber\\
	&&~~~~~~~~~~~~+Re\{(C_{RL}+C_{RR})(C_{RLLR}^*+C_{RLRL}^*)\})\nonumber\\
	&&-2m_sm_l(Re\{(C_{LL}+C_{LR})(C_{RLLR}^*+C_{RLRL}^*)\}\nonumber\\
	&&~~~~~~~~~~~~+Re\{(C_{RL}+C_{RR})(C_{LRLR}^*+C_{LRRL}^*)\})\nonumber\\
	&&+24(m_b+m_s)m_lRe\{(C_{LL}-C_{LR}+C_{RL}-C_{RR})C_T^*\}\nonumber\\
	&&+48(-m_b+m_s)m_lRe\{(C_{LL}-C_{LR}+C_{RL}-C_{RR})C_{TE}^*\}].\label{eqn:unFBmassive}
\end{eqnarray}
If $C_{BR}=C_{SL}=-2C_7^{eff}$, this corresponds with Eq. (A6) in Ref. \cite{FKMY}.


\end{document}